\documentclass[
 reprint,
 amsmath,amssymb,
 aps,
 footinbib,
 superscriptaddress
]{revtex4-2}

\usepackage{placeins}
\usepackage{lipsum}
\usepackage{graphicx}
\usepackage{epstopdf}

\usepackage{dcolumn}
\usepackage{bm}
\usepackage[colorlinks=true, citecolor={blue!80!black}, urlcolor={blue!50!black}, linkcolor = {blue!80!black}]{hyperref}
\usepackage[colorinlistoftodos]{todonotes}
\usepackage[capitalize]{cleveref}
\usepackage[english]{babel}
\usepackage{braket}
\usepackage[utf8]{inputenc}
\usepackage{physics}

\usepackage{siunitx}
\DeclareSIUnit{\belmilliwatt}{Bm}
\DeclareSIUnit{\dBm}{\deci\belmilliwatt}

\newcommand{\Vc}{$V_\mathrm{C}$}

\begin{document}

\preprint{APS/123-QED}

\title{Strong tunable coupling between two
distant superconducting spin qubits}

\author{Marta Pita-Vidal}
\thanks{These two authors contributed equally.}
\affiliation{QuTech and Kavli Institute of Nanoscience, Delft University of Technology, 2600 GA Delft, The Netherlands}

\author{Jaap J. Wesdorp}
\thanks{These two authors contributed equally.}
\affiliation{QuTech and Kavli Institute of Nanoscience, Delft University of Technology, 2600 GA Delft, The Netherlands}

\author{Lukas J. Splitthoff}
\affiliation{QuTech and Kavli Institute of Nanoscience, Delft University of Technology, 2600 GA Delft, The Netherlands}

\author{Arno Bargerbos}
\affiliation{QuTech and Kavli Institute of Nanoscience, Delft University of Technology, 2600 GA Delft, The Netherlands}

\author{Yu Liu}
\affiliation{Center for Quantum Devices, Niels Bohr Institute, University of Copenhagen, 2100 Copenhagen, Denmark}

\author{Leo P. Kouwenhoven}
\affiliation{QuTech and Kavli Institute of Nanoscience, Delft University of Technology, 2600 GA Delft, The Netherlands}

\author{Christian Kraglund Andersen}
\affiliation{QuTech and Kavli Institute of Nanoscience, Delft University of Technology, 2600 GA Delft, The Netherlands}

\date{July 28, 2023}

\begin{abstract}

Superconducting (or Andreev) spin qubits have recently emerged as an alternative qubit platform with realizations in semiconductor-superconductor hybrid nanowires~\cite{Hays2021, PitaVidal2023}. 
In these qubits, the spin degree of freedom is intrinsically coupled to the supercurrent across a Josephson junction via the spin-orbit interaction, which facilitates fast, high-fidelity spin readout using circuit quantum electrodynamics techniques~\cite{Hays2020}.
Moreover, this spin-supercurrent coupling has been predicted to facilitate inductive multi-qubit coupling~\cite{Chtchelkatchev2003,
Padurariu2010}.
In this work, we demonstrate a strong supercurrent-mediated coupling between two distant Andreev spin qubits.
This qubit-qubit interaction is of the longitudinal type and we show that it is both gate- and flux-tunable up to a coupling strength of~\SI{178}{MHz}. Finally, we find that the coupling can be switched off in-situ using a magnetic flux.
Our results demonstrate that integrating microscopic spin states into a superconducting qubit architecture can combine the advantages of both semiconductors and superconducting circuits and pave the way to fast two-qubit gates between remote spins. 
\end{abstract}

\maketitle

Semiconducting spin qubits~\cite{diVincenzo1998, Hanson2007} have proven to be a promising platform for quantum information processing. In such qubits, quantum information is encoded in the spin degree of freedom of electrons or holes localized in quantum dots, which leads to long lifetimes and a naturally large energy separation between computational and non-computational states. 
Moreover, their small size makes them attractive candidates for large-scale quantum devices~\cite{Vandersypen2017, Burkard2023}. 
However, it remains challenging to engineer a direct spin-spin coupling between remote spin-qubits as their interaction strength decays rapidly with distance. Ongoing efforts to overcome this challenge focus on engineering a coupling between distant spin-qubits mediated by microwave photons in superconducting resonators~\cite{Mi2018, Samkharadze2018, Landig2018, Borjans2020, HarveyCollard2022, Yu2023}. 
For such photon-mediated spin-spin coupling, the interaction strength is currently limited to the order of \SI{10}{MHz}, which makes the implementation of fast, long-range two-qubit gates an outstanding challenge~\cite{Burkard2023, HarveyCollard2022}. Moreover, the transverse character of the coupling puts a constraint on the available qubit frequencies.

\begin{figure}[h!]
\includegraphics{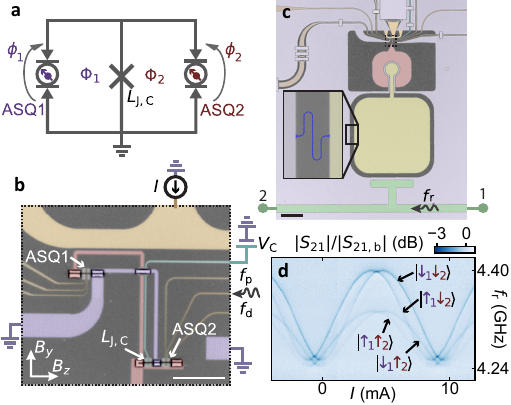}
\caption{ {\bf Device and readout.} 
{\bf a} Circuit diagram of two coupled Andreev spin qubits (ASQ1 and ASQ2) connected to a coupling junction with a tunable Josephson inductance $L_{\rm J, C}$. $\Phi_{1}$ and $\Phi_{2}$ are the magnetic fluxes through the two loops.
{\bf b} False-colored optical microscope image of the device. The ASQs are placed between a transmon island (red) and ground (purple). The three Josephson junctions are implemented in two separate Al/InAs nanowires, with one containing ASQ1 and the other containing ASQ2 and the coupling junction. The in-plane magnetic field directions are denoted as $B_z$ and $B_y$, approximately parallel and perpendicular to the nanowires axis, respectively. Additional flux control is achieved through the applied current $I$ in the flux line (amber). Each ASQ is electrostatically controlled by three gates below the nanowire (brown), while the coupling junction is controlled by one gate line (cyan) at voltage $V_{\rm C}$. The drive tones $f_{\rm d}$ and $f_{\rm p}$ are applied through the central gate of ASQ2. See the Supplementary Information~\cite{Supplement} for further details about the geometry of the loops area.
{\bf c} Zoomed out false-colored optical microscope image showing the transmon island (red) capacitively coupled to a lumped-element readout resonator, consisting of a capacitor (yellow) and an inductor (blue, inset). The resonator is further capacitively coupled to a coplanar waveguide (green center conductor) with input and output ports labeled as 1 and 2, respectively. A readout tone $f_{\rm r}$ is applied through the waveguide. Scale bars in {\bf b} and {\bf c} correspond to \SI{10}{\micro m} and \SI{100}{\micro m}, respectively.
{\bf d} Amplitude of the transmission through the readout circuit, $|S_{21}|$, divided by the background, $|S_{21, \rm b}|$, as a function of the current through the flux line, $I$. The measurement is performed at a magnetic field of $B_z=0$ with a fixed $\Phi_{2}\sim -\Phi_0/4$, set by $B_y=$~\SI{-1.04}{mT}.
}
\label{fig:fig1}
\end{figure}

An alternative approach to engineer remote spin-spin coupling is to embed the spin-qubit into a Josephson junction creating a so-called Andreev spin qubit (ASQ)~\cite{Hays2021, PitaVidal2023}, where the qubit states carry a spin-dependent supercurrent~\cite{Tosi2018, Hays2020, Wesdorp2021,Hays2021, Wesdorp2022, Bargerbos2022b, PitaVidal2023}.
Recent experiments have demonstrated that a single ASQ can be operated coherently with strong coupling of the spin states to superconducting circuits~\cite{Hays2021, PitaVidal2023}. Similarly, it has been predicted that large spin-dependent supercurrents can lead to strong, longitudinal, long-range and tunable spin-spin coupling~\cite{Chtchelkatchev2003, Padurariu2010}, thus, overcoming the challenges imposed by the coupling being only a second-order interaction in previous photon-mediated implementations of spin-spin coupling as well as circumventing any strong constraints on the qubit frequencies.

Here, we investigate the supercurrent-mediated coupling between two ASQs by analyzing the influence of a shared Josephson inductance on the coupling strength using the setup in Fig.~\ref{fig:fig1}. Specifically, we design a device formed by two Andreev spin qubits, ASQ1 and ASQ2, connected in parallel to a third Josephson junction with gate-tunable Josephson inductance, thus defining two superconducting loops (Fig.~\ref{fig:fig1}a).
Microscopically, the longitudinal coupling between the qubits directly results from the main characteristic of Andreev spin qubits: their spin to supercurrent coupling. The phase-dependent frequency of one qubit results in a spin-state-dependent circulating supercurrent through the loop arm containing the other qubit.
We show that the qubit-qubit coupling in this configuration can be in-situ controlled by the flux through the superconducting loops as well as by changing the Josephson inductance of the shared junction using an electrostatic gate. 
In particular, we reach the strong longitudinal coupling regime where the coupling strength is larger than the qubit linewidths.
Moreover, we show that the coupling can be switched fully off for particular values of the flux, which makes this platform appealing as an alternative for implementing fast flux-controlled two-qubit gates between spin qubits.

\section{Device}
In our device, each ASQ is hosted in a quantum dot Josephson junction which is implemented in a separate Al/InAs nanowire and controlled by three electrostatic gates placed beneath the nanowires (Fig.~\ref{fig:fig1}b). Throughout this work, the gate voltages are fixed as specified in the Supplementary Information~\cite{Supplement}. Moreover, we define an additional regular Josephson junction with gate-tunable Josephson inductance $L_{\rm J, C}$ in one of the nanowires. The nanowires are galvanically connected to a NbTiN circuit which defines the superconducting loops forming a double-loop superconducting quantum interference device (SQUID). We denote by $\Phi_1$ and $\Phi_2$ the external magnetic fluxes through each of the loops. 
The qubit frequency for ASQ$i$, $f_i$, where $i=1,2$, is set by the energy difference between the spin-states,  $\ket{\uparrow_i}$ and $\ket{\downarrow_i}$, which is controlled by the magnetic field due to the Zeeman effect. We denote the in-plane magnetic field directions as $B_z$, approximately along the nanowires, and $B_y$, approximately perpendicular to the nanowires. See also Supplementary Information \cite{Supplement} for additional details on the field alignment.
The $B_y$ component of the magnetic field is moreover used to tune $\Phi_{\rm 1}$ and $\Phi_{\rm 2}$. Note that, while $B_y$ is applied in the chip plane, it still threads flux through the loops due to the elevation of the nanowires with respect to the NbTiN circuitry. This reduces flux jumps compared to using out-of-plane field $B_x$ for flux tuning, as discussed in Ref.~\cite{Wesdorp2022}.  $\Phi_{\rm 1}$ and $\Phi_{\rm 2}$ set the phase drops over the junctions, $\phi_1 \sim \frac{2\pi}{\Phi_0}\Phi_1$ and $\phi_2 \sim \frac{2\pi}{\Phi_0}\Phi_2$ in the limit of small $L_{\rm J, C}$, where  $\Phi_0$ denotes the magnetic flux quantum. The current through the flux line, $I$, tunes $\Phi_{\rm 1}$ and leaves $\Phi_{\rm 2}$ nearly unaffected, as the loop corresponding to $\Phi_{\rm 2}$ is placed near the symmetry axis of the flux line (see Supplementary information~\cite{Supplement}). The drive pulses, with frequencies $f_{\rm d}$ and $f_{\rm p}$, are sent through the central gate of ASQ2 and are used to drive both qubits. We find that it is possible to drive ASQ1 using the gate line of ASQ2 possibly due to cross-coupling between the gate lines corresponding to both qubits or to cross-coupling between the gate line and the transmon island.
The coupling junction is controlled by a single electrostatic gate whose voltage, \Vc,  is varied to tune $L_{\rm J, C}$~\cite{doh_tunable_2005}.

To enable readout of the ASQ states, the double-loop SQUID in which the ASQs are hosted is placed between a superconducting island (red) and ground (purple), forming a transmon circuit~\cite{Koch2007, Larsen2015, deLange2015} (Fig.~\ref{fig:fig1}b,~c).
These circuit elements are implemented in \SI{20}{nm}-thick NbTiN for magnetic field compatibility~\cite{Samkharadze2016, Kroll2018, Kroll2019, PitaVidal2020, Kringhoj2021, Uilhoorn2021, Wesdorp2022}. The transmon frequency depends on the energy-phase relation of the double-loop SQUID, which in turn depends on the states of both ASQs~\cite{Bargerbos2022b}. 
The transmon is subsequently dispersively coupled to a lumped element readout resonator, which is coupled to a feedline implemented with a coplanar waveguide and monitored in transmission using a probe tone at frequency $f_{\rm r}$. The readout mechanism is illustrated in Fig.~\ref{fig:fig1}d, which shows the four possible frequencies of the readout resonator, caused by the different dispersive shifts of the four spin states of the combined ASQ1-ASQ2 system~\cite{Blais2004}: $\left\{\ket{\uparrow_1\uparrow_2}, \ket{\uparrow_1\downarrow_2},\ket{\downarrow_1,\uparrow_2},\ket{\downarrow_1\downarrow_2}\right\}$. Note that spin is not a well defined quantum number for these states, see~\footnote{In an ASQ, the spin is hybridized with spatial degrees of freedom, and thus the eigenstates are rather pseudo-spin states. Similar to previous works~\cite{Hays2021, PitaVidal2023}, we will refer to the eigenstates as spins for simplicity.}. The measurement is taken at zero magnetic field where all spin states are thermally occupied on average, since the energy splitting between them is between 0.5 and $\SI{1}{\giga\hertz}$ (see Supplementary Information~\cite{Supplement}), which is smaller than typical effective temperatures on the order of $\SI{100}{\milli\kelvin}$ observed in these devices~\cite{PitaVidal2023}. Therefore, the lines corresponding to all four states are visible. This result already illustrates the presence of two separate ASQs in the system. We will now move on to the characterization of these qubits before we turn our attention to the two-qubit coupling.

\section{Individual Andreev spin qubit characterization}
\begin{figure}[t]
\includegraphics{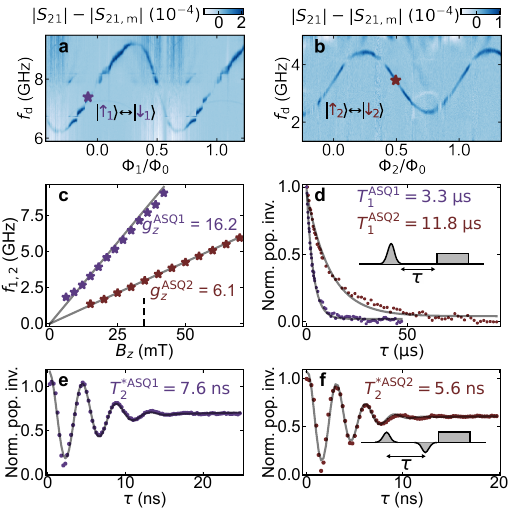}
\caption{ {\bf Individual Andreev spin qubit properties}. \textbf{a, b} Readout signal amplitude with the median background subtracted, $|S_{21}| - |{S_{21, {\rm m}}}|$, showing qubit spectroscopy of ASQ1 (versus $\Phi_1$) and ASQ2 (versus $\Phi_2$), respectively. During spectroscopy of one qubit, the other qubit is turned off by setting its gates to \SI{-1}{\volt}. We set $B_r=$~\SI{35}{mT} for both panels (indicated in {\bf c} with a dashed line).
\textbf{c} Qubit frequency versus $B_z$ for both ASQs. $f_i$ is calculated as the average between its maximum and minimum values versus flux. The grey lines indicate a linear fit to the data from which we extract the $g$-factors indicated in the labels.
\textbf{d} Energy decay time ($T_1$) measurements of both ASQs at the frequency setpoints indicated in {\bf a}, {\bf b}  ($f_1$~=~\SI{7.4}{\giga\hertz} and $f_{2}$~=~\SI{3.4}{\giga\hertz}, respectively). The experiment was performed by sending a $\pi$-pulse followed, after a delay $\tau$, by a readout pulse (see inset). 
\textbf{e}, \textbf{f} Measurements of the coherence times ($T_2^*$) of ASQ1 and ASQ2 at the same setpoints, measured using a Ramsey experiment. Oscillations with a period of \SI{4}{\nano\second} (for {\bf e}) and \SI{3}{\nano\second} (for {\bf f}) 
are realized by adding a phase to the final $\pi/2$ pulse proportional to the delay time $\tau$. The pulse sequence is shown in the inset for a phase of $\pi$. $T_2^*$ is extracted by fitting a sine with a Gaussian decay envelope. The experiments were performed using Gaussian pulses with a FWHM of $\SI{4}{\nano\second}$. All datasets are averaged over $3\cdot 10^5$ shots, readout time ranges from 1 to \SI{2}{\micro \second} and the total measurment time for $T_2^{*,{\rm ASQ}i}$ ranges from around 10~min for ASQ1 to around 30~min for ASQ2. The normalized population inversion on the y-axis of panels {\bf d}-{\bf f} is defined as the measured signal normalized by the signal difference between having sent no pulse and a $\pi$-pulse before the readout pulse.}
\label{fig:fig2}
\end{figure}

We first characterize each ASQ separately, while the junction containing the other qubit is pinched-off electrostatically using the voltages on its gates (\cref{fig:fig2}), following the methods of Ref.~\cite{PitaVidal2023}.
To set the qubit frequencies, we apply a magnetic field $B_r=$~\SI{35}{mT} in the $y$-$z$ plane, 0.1 radians away from the $B_z$ direction (see Supplementary Information~\cite{Supplement}). This field sets  $f_1\in$~[6,~9]~GHz  and $f_2\in$~[2,~4.5]~GHz for ASQ1 and ASQ2, respectively. We note that the qubit frequencies are significantly different due to mesoscopic fluctuations in the gate-dependence of the spin-orbit direction and $g$-factor of each ASQ, see also Fig.~\ref{fig:fig2} and Supplementary Information. Qubit spectroscopy is then performed by monitoring the transmission through the feedline near the readout-resonator frequency, while applying a drive tone with frequency $f_{\rm d}$ to the central gate line of ASQ2, see Fig.~\ref{fig:fig2}a,~b. 
On resonance with the qubit transition, we observe a strong change in transmission because spin-orbit coupling and a magnetic field enable electrical driving of the spin~\cite{Metzger2021, Wesdorp2022, Bargerbos2022b}. 
The qubit frequencies, $f_1$ and $f_2$, can be tuned by flux, as shown in Fig.~\ref{fig:fig2}a and b.
Note that the phase dispersion is expected to be sinusoidal, see Ref.~\cite{Padurariu2010, pavevsic2022b}, as is the case for ASQ2. However, for ASQ1 we rather observe a skewed sine. From the ratio of the inductance of ASQ1 and $L_{\rm J,C}$ we rule out a non-linear flux-phase relation, so the skewness is currently of unknown origin and could be related to higher orbitals in the quantum dot.
While flux tuning provides fine-tuning of the qubit frequency within a frequency band of a few GHz set by the spin-orbit coupling strength, we can also tune the qubit frequencies over a larger range by varying the magnetic field, due to the Zeeman effect. From the magnetic field dependence of the frequencies we extract the $g$-factor of each ASQ, see Fig.~\ref{fig:fig2}c. We find that the different $g$-factors are consistent with earlier work~\cite{Vaitekenas2018, Bargerbos2022b, Wesdorp2022}, see also Supplementary Information~\cite{Supplement}. 

Next, we characterize the coherence properties of each ASQ at the frequencies indicated with markers in Fig.~\ref{fig:fig2}a and b.
At these setpoints, we extract energy decay times of $T_1^{{\rm ASQ}1}=3.3\pm\SI{0.1}{\micro s}$ and $T_1^{{\rm ASQ}2}=11.8\pm\SI{0.4}{\micro s}$ for ASQ1 and ASQ2, respectively, where the reported uncertainties are the $1\sigma$ confidence intervals from the fit. These decay times are to a large extent limited by Purcell decay to the transmon qubit (see Supplementary Information \cite{Supplement}).
Furthermore, from a Ramsey experiment, we extract dephasing times of $T_2^{* {\rm ASQ}1}=7.6$ $\pm $~\SI{0.2}{n s} and $T_2^{* {\rm ASQ}2}=5.6$ $ \pm $~\SI{0.2}{n s} for ASQ1 and ASQ2, respectively, which are comparable to times found in earlier works~\cite{Hays2021, PitaVidal2023}. For these measurements, we use Gaussian pulses with a full width at half-maximum (FWHM) of \SI{4}{\nano\second}, which is comparable to $T_2^*$. Therefore, the $\pi/2$ pulses cannot be considered instantaneous, which is the conventional assumption in a Ramsey experiment. Rather, a non-zero overlap of the pulses of order $T_2^*$ can result in an overestimation of the extracted $T_2^*$, as further discussed in the Supplementary Information~\cite{Supplement}.
Therefore, these numbers should be interpreted as an upper bound to the pure dephasing times.
Furthermore, we extract echo times of $T_{2 \rm E}^{{\rm ASQ}1}=$~17.3~$\pm$~\SI{0.4}{n s} and $T_{2 \rm E}^{{\rm ASQ}2}=17.4$~$\pm$~\SI{0.4}{n s}, see Supplementary Information \cite{Supplement}, three times larger than $T_2^{*}$, which points at low-frequency noise being a strong contributor to dephasing, consistent with previous observations in InAs-based spin qubits~\cite{NadjPerge2010,Hays2021, PitaVidal2023}.

\section{Longitudinal coupling}

Having two Andreev spin qubits, we describe the joint system by the following Hamiltonian with the two qubits coupled longitudinally with coupling strength $J$~\cite{Padurariu2010}:
\begin{equation}
  H = -\frac{\hbar\omega_1}{2}\sigma_{1}^z -\frac{\hbar\omega_2}{2}\sigma_{2}^z - \frac{hJ}{2}\sigma_1^z\sigma^z_2,
  \label{eq:coupling_Hamiltonian}
\end{equation}
where $\omega_i = 2\pi f_i$ and $\sigma_i^z = \ket{\downarrow_i}\bra{\downarrow_i} - \ket{\uparrow_i}\bra{\uparrow_i}$ denote the phase-dependent spin-flip frequency and the $z$ Pauli matrix of ASQ$i$, respectively, $h$ is the Planck constant and $\hbar=h/(2\pi)$. 
In this description, the longitudinal term $\frac{-hJ}{2}\sigma_1^z\sigma^z_2$ originates from the fact that the spin-dependent supercurrent of ASQ1 induces a spin-dependent phase difference over ASQ2, thus changing its transition frequency by $\pm J$, and vice versa.
Importantly, the longitudinal coupling does not arise from direct wavefunction overlap~\cite{Spethmann2022} or magnetic interactions as the spins are separated by a distance of approximately $ \SI{25}{\micro\meter}$.
From this physical understanding of the interaction, we can express the coupling strength $J$ as a function of the circuit parameters by~\cite{Padurariu2010}
\begin{equation}
J(L_{\rm J, C}, \Phi_1, \Phi_2)  = \frac{1}{2h} \frac{L_{\rm J, C} L_{\rm ASQ}(\Phi_1, \Phi_2)}{L_{\rm J, C}+L_{\rm ASQ}(\Phi_1, \Phi_2)} I_{1}(\Phi_1) I_{2}(\Phi_2).
\label{eq:coupling_energy}
\end{equation}
Here, we define $L_{\rm ASQ}(\Phi_1, \Phi_2)$ as the total spin-independent inductance of the two ASQs in parallel and the magnitude of the spin-dependent current is captured by $I_{i}(\Phi_i)$ which denotes the difference in supercurrent across ASQ$i$ for its two possible spin states.
In this expression, one of the main features of the device becomes apparent: the coupling is tunable with flux and can be switched to zero when either $I_{1}$ or $I_{2}$ are set to zero.

\begin{figure}[h!]
\includegraphics{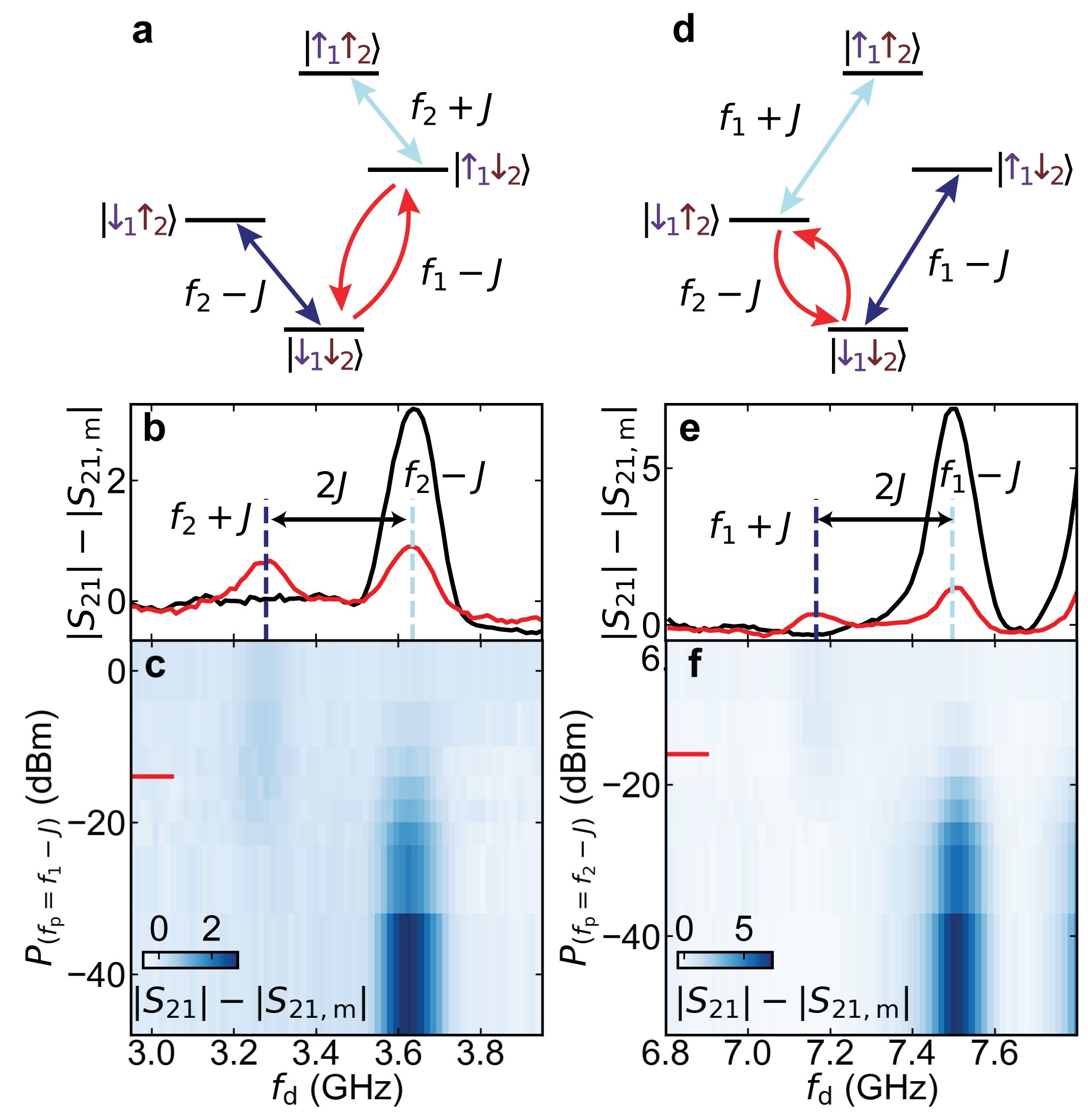}
\caption{\label{fig:fig3} {\bf Strong longitudinal coupling between the two Andreev spin qubits.} 
{\bf a} Energy level diagram of the combined ASQ1-ASQ2 system with the levels (horizontal black lines) labeled by the states of both qubits (ASQ1 in purple, ASQ2 in maroon). The diagonal arrows denote the two different transition frequencies ($f_2 \pm J$) of ASQ2 depending on the state of ASQ1. Note that $J$ is negative in this illustration and for the data presented in this figure. 
{\bf b} Spectroscopy of ASQ2 as a function of the drive frequency $f_{\rm d}$. The black and red lines indicate the readout signal amplitude with the background subtracted, $|S_{21}| - |{S_{21, {\rm m}}}|$, with and without a pump tone resonant with ASQ1 at frequency $f_{\rm p} = f_1-J$, respectively. The pump tone is indicated with red arrows in {\bf a}. 
{\bf c} Power dependence of the pump tone. The red line indicates the power used for the red line in {\bf c}. We indicate the power at the source output.
{\bf d}-{\bf f} Similar to panels {\bf a}-{\bf c}, but with the roles of ASQ1 and ASQ2 exchanged. In this case, the pump tone drives ASQ2 at a frequency $f_{\rm p}=f_2-J$, while performing spectroscopy of ASQ1. 
}
\end{figure}

We now proceed to investigate the spin-spin coupling at the same gate voltages and magnetic field used for~\cref{fig:fig2}. To this end, we open both loops simultaneously and set $\Phi_1$ and $\Phi_2$ at points where the slopes of the qubit frequencies $\partial f_i/\partial \Phi_i\propto I_{i}$ are large, close to $\Phi_1\sim0$ and $\Phi_2\sim\Phi_0/2$. 
When the two qubits are longitudinally coupled, the transition frequency of each of them depends on the state of the other, as schematically depicted in~\cref{fig:fig3}a and d.
In each panel, the blue arrows indicate the two possible frequencies of one qubit, separated by twice the coupling strength, $J$, for the two possible states of the other qubit. To determine the magnitude of the coupling strength, we perform the following measurements:
First, we determine $f_2-J$ by performing qubit spectroscopy of ASQ2 starting from the ground state, $\ket{\downarrow_1\downarrow_2}$, where ASQ1 is in the spin-down state (black trace in Fig.~\ref{fig:fig3}b).
Then, we repeat the spectroscopy while applying another continuous pump tone at a frequency $f_{\rm p}$ resonant with the spin-flip transition of ASQ1, driving $\ket{\downarrow_1\downarrow_2}\leftrightarrow\ket{\uparrow_1\downarrow_2}$.
The presence of this additional tone results in ASQ1 being in a mixture of $\ket{\downarrow_1}$ and $\ket{\uparrow_1}$.
When performing spectroscopy of ASQ2 under these conditions (red trace in Fig.~\ref{fig:fig3}b), we observe the emergence of a second peak corresponding to the shifted frequency of ASQ2 due to ASQ1 having population in its excited state, $\ket{\uparrow_1}$. This frequency splitting arises from the longitudinal coupling term and, thus, we determine the value of $J=-178 \pm 3$~MHz from a double Gaussian fit as half of the difference between the two frequencies (see the Supplementary Information for details on the fit procedure \cite{Supplement}).
Since the coupling term is symmetric with respect to the two qubits, we should observe the same frequency splitting when we exchange the roles of ASQ1 and ASQ2, see Fig.~\ref{fig:fig3}e (note that the increase in amplitude around \SI{7.8}{GHz} is unrelated to the ASQs but due to a resonance of the traveling wave parametric amplifier). From this measurement, we extract a value of $J=-165 \pm 4$~MHz similar to the value we extracted before. We speculate that the modest difference between the values of $J$ extracted from the measurements of both qubits may be due to temporal instabilities, which we found to be present in the system.
We additionally measure the qubit spectroscopy as a function of the pump tone power, shown in Fig.~\ref{fig:fig3}c and f, and we observe a power dependence on the peak amplitude. At low powers, not enough excited population is generated in the ASQ while the second peak gradually appears at higher powers. At too high powers, the readout resonator shifts too much due to the non-linearity of the resonator mode and it becomes more lossy, which results in a reduced signal (at even higher power both peaks fully disappear). Additional data and a numerical analysis of the expected pump power dependence and relative peak heights, in agreement with the experimental observations, can be found in the Supplementary Information \cite{Supplement}.

Next, we compare the extracted value of $J$ to the linewidth of the ASQ transitions and find 
$J =$~\SI{165}{MHz}~$>$~\SI{28}{MHz}~$= 1/(2\pi T_2^{* {\rm ASQ}2})$, 
indicating that the system is in the strong longitudinal coupling regime. 
This value of $J$ puts a speed limit for a controlled-Z two-qubit gate at a time of $t=1/(4J)=\SI{1.4}{\nano\second}$ and a coherence limit on the average gate fidelity of around $85\%$, which will be explored in future experiments. Such a two-qubit gate, combined with single qubit rotations, enables a universal set of gates. On the other hand, such a fast gate would require distortion-free flux pulses~\cite{Rol2020}, with a rise time much smaller than the gate time of \SI{1.5}{\nano\second}. This two-qubit gate time is much faster than typical fast two-qubit gates with superconducting qubits ($10-\SI{45}{\nano\second}$~\cite{Arute2019, Rol2019}) and comparable to the fastest short distance exchange gates in spin qubits coupled via directly overlapping wavefunctions~\cite{diVincenzo1998, he_two-qubit_2019, hendrickx_four-qubit_2021}.

\section{Tunability of the coupling strength}

\begin{figure}[h!]
\includegraphics{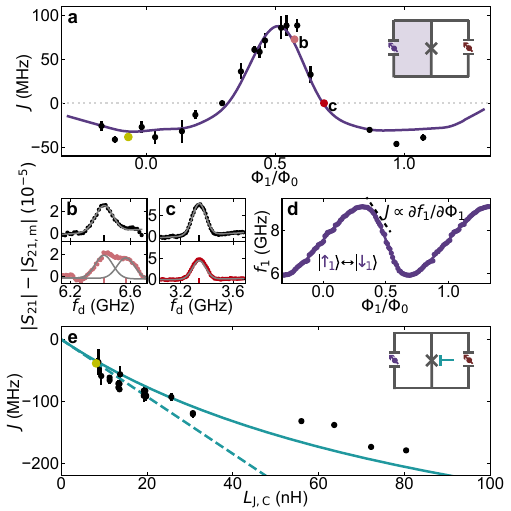}
\caption{\label{fig:fig4}{\bf Tunability of the coupling strength.} 
{\bf a} Qubit-qubit coupling strength, $J$, as a function of flux in the loop containing ASQ1, $\Phi_1$, see also inset, at fixed $\Phi_2\sim0.51\Phi_0$.  
The markers and error bars represent the best-fit values of $J$ (see panels {\bf b, c}) and their estimated standard errors (one-sigma confidence intervals), respectively. The purple line shows the expected dependence from Eq.~\eqref{eq:coupling_energy}.
{\bf b, c} Representative fits at two $\Phi_1$ points highlighted with colored (and letter marked) markers in {\bf a}. The signal measured in the absence of a pump tone (black markers) is fit with a single Gaussian (black line), to determine $f_i-J$ (vertical black line in the x axis). The signal measured in the presence of a pump tone at the other ASQ (colored markers) is additionally fitted (colored line) to determine $f_i+J$. The grey lines in {\bf b} show the two individual Gaussians. 
{\bf d} Frequency of ASQ1, $f_{1}$, versus $\Phi_1$ (markers) and interpolation (line) used to estimate $I_{1}(\Phi_1)\sim h \partial{f_{1}}/\partial{\Phi_1}$. 
{\bf e} Qubit-qubit coupling strength $J$ at fixed $\Phi_1=-0.07 \Phi_0$ and as a function of $L_{\rm J, C}$, which is varied using the gate-voltage at the coupling junction (see inset). The continuous line shows the dependence from Eq.~\eqref{eq:coupling_energy}, 
while the dashed line shows a linear dependence $Jh= L_{\rm J, C} I_{1} I_{2}/2$.
The yellow marker in {\bf a} and in {\bf e} is a shared point between the two panels. 
 }
\end{figure}

We have so far investigated the coupling strength at fixed gate voltages and flux. We now investigate the dependence of $J$ on different control parameters and demonstrate that it is tunable as predicted by~Eq.~\eqref{eq:coupling_energy}~\cite{Padurariu2010}. We vary $\Phi_1$ using the flux line, see Fig.~\ref{fig:fig4}a, and find that the coupling strength is directly proportional to $I_{1}$, as expected. The current difference across ASQ1, $I_1$, is extracted from a measurement of the qubit frequency as a function of flux, as shown in Fig.~\ref{fig:fig4}d.
Note that, by varying the flux, we not only vary the magnitude of $J$, but also switch its sign, crossing zero coupling.
Thus, the two ASQs can be fully uncoupled by setting $J=0$ at the flux points which maximize or minimize $f_i(\Phi_i)$, and where thus $I_i=0$, for either one of the qubits.
The coinciding of zero coupling with these frequency-extrema is useful as these are the first-order flux-insensitive points of the qubit transition frequency. 
Two representative situations in which the ASQs are coupled and uncoupled at nearby flux points are shown in Fig.~\ref{fig:fig4}b and c, respectively. The data was measured and analyzed using the same procedure as described for~\cref{fig:fig3}.

We overlay the $\Phi_1$-dependence of the coupling strength with the expected dependence from~\cref{eq:coupling_energy}.
The values of $L_{\rm J, C}=$~\SI{8.4}{nH} and $I_{2}\sim h\partial f_2/\partial \Phi_2|_{\Phi_2=0.51\Phi_0} =$~\SI{-2.52}{nA} are fixed and independently extracted from measurements of the transmon frequency and of $f_2(\Phi_2)$, respectively. $L_{\rm ASQ}(\Phi_1)$ is calculated as the parallel combination of the spin-independent Josephson inductances of both qubits, which are determined from separate transmon spectroscopy measurements (see Supplementary Information \cite{Supplement}) and 
$I_{1}(\Phi_1)\sim h \partial{f_{1}}/\partial{\Phi_1}$ is estimated from Fig.~\ref{fig:fig4}d.  
As shown in Fig.~\ref{fig:fig4}a, the measured $J(\Phi_1)$ is in good agreement with Eq.~\eqref{eq:coupling_energy}.

Finally, we investigate the $L_{\rm J, C}$ tunability of $J$ by fixing  $\Phi_1=-0.07 \Phi_0$, which sets $I_{\rm 1}=$~\SI{2.16}{nA}, and varying the value of \Vc~(see Supplementary Information for the corresponding qubit parameters~\cite{Supplement}).
We observe an increase of the magnitude of $J$ as the value of $L_{\rm J, C}$ is increased, as shown in Fig.~\ref{fig:fig4}e. 
The measured data follows to a large extent the dependence expected from Eq.~\eqref{eq:coupling_energy}, indicated with a continuous line in Fig.~\ref{fig:fig4}e.
The $|J|$ increase is limited to a maximum when the coupling junction $L_{\rm J, C}$ becomes comparable to the finite spin-independent inductance $L_{\rm ASQ}$ of the ASQs. For the solid line in \cref{fig:fig4}e we use the independently measured value $L_{\rm ASQ}(\Phi_1=-0.07 \Phi_0, \Phi_2=0.51\Phi_0) =$~\SI{102.0}{nH}. For comparison, the dashed line depicts the limit of $L_{\rm ASQ}\gg L_{\rm J,C}$.

\section{Conclusions}\label{sec:conclusions}

In conclusion, we have extended earlier results demonstrating single Andreev spin qubits~\cite{Hays2021, PitaVidal2023} and integrated two InAs/Al-based ASQs within a single transmon circuit. The two ASQs are separated by around \SI{25}{\micro m}, two orders of magnitude larger than the size of the individual qubit wavefunctions. Both ASQs showed comparable coherence properties to those reported in prior work~\cite{Hays2021,PitaVidal2023}.  
We have shown strong supercurrent-mediated coupling between the two Andreev spin qubits and found that the coupling strength, $J$, can be tuned with either a magnetic flux or an electrical voltage.
In particular, we have shown that $J$ can be fully suppressed using a magnetic flux. This switchability of the coupling is essential for the use of longitudinally coupled Andreev spin qubits to perform quantum computation. Furthermore, the high sign and magnitude tunability of $J$ could have applications for the use of Andreev spin qubits to perform analog quantum simulations. More generally, Andreev spin qubits could in the future provide an independent platform for quantum computing and simulation or, alternatively, they may be incorporated into existing spin qubit platforms and serve as readout modules or long-distance couplers. Independently of the precise use-case for Andreev spin qubits, we emphasize that strong spin-spin coupling as demonstrated here will be an essential requirement, although smaller dephasing rates would be desired.

Previous works suggest that one possible mechanism limiting dephasing is coupling to the large nuclear spins of InAs~\cite{NadjPerge2010, Hays2021, PitaVidal2023}. While the origin of dephasing must be further investigated, this suggests that a possible route to increase the dephasing times is implementing Andreev spin qubits in an alternative nuclear-spin-free material such as germanium  \cite{Hendrickx2018, Scappucci2021, Tosato2023, Valentini2023}.
We expect that future efforts using alternative materials could both provide a path towards integration in more established semiconductor-based quantum architectures as well as strongly increased coherence times. If longer coherence times can be achieved, in combination with the strong qubit-qubit coupling demonstrated here, Andreev spin qubits will emerge as an encouraging platform for the realization of high-fidelity two-qubit gates between remote spins.

\begin{acknowledgments}
We thank B. van Heck, A. Kou, G. de Lange, V. Fatemi, P. Kurilovich, S. Diamond, T. Connolly, H. Nho, C. Boettcher, V. Kurilovich and X. Xue for discussions and their feedback on this manuscript. We thank Y. Nazarov for insightful discussions.
We further thank Peter Krogstrup for guidance in the material growth. 
This work is part of the research project ‘Scalable circuits of Majorana qubits with topological protection’ (i39, SCMQ) with project number 14SCMQ02, which is (partly) financed by the Dutch Research Council (NWO). It has further been supported by the Microsoft Quantum initiative. C.K.A. acknowledges support from the Dutch Research Council (NWO).
\end{acknowledgments}

\section*{Data availability}
Data as well as processing and plotting scripts are available online at \url{https://doi.org/10.4121/e10185d0-026e-480f-bbaa-3448c6e1b9a2}.

\section*{Author contributions}
J.J.W., M.P.V., and C.K.A. conceived the experiment.
Y.L. developed and provided the nanowire materials.
J.J.W, M.P.V., L.S. and A.B. prepared the experimental setup and data acquisition tools.
L.S. deposited the nanowires.
J.J.W, M.P.V. and A.B designed the device.
J.J.W and M.P.V. fabricated the device, performed the measurements and analysed the data, with continuous feedback from L.S., A.B. and C.K.A.
L.P.K. and  C.K.A. supervised the work.
J.J.W., M.P.V., and C.K.A. wrote the manuscript with feedback from all authors.

\bibliography{bibliography.bib}

\end{document}


\preprint{APS/123-QED}

\title{Supplementary information: Strong tunable coupling between two distant superconducting spin qubits}

\author{Marta Pita-Vidal}
\thanks{These two authors contributed equally.}
\affiliation{QuTech and Kavli Institute of Nanoscience, Delft University of Technology, 2600 GA Delft, The Netherlands}

\author{Jaap J. Wesdorp}
\thanks{These two authors contributed equally.}
\affiliation{QuTech and Kavli Institute of Nanoscience, Delft University of Technology, 2600 GA Delft, The Netherlands}

\author{Lukas J. Splitthoff}
\affiliation{QuTech and Kavli Institute of Nanoscience, Delft University of Technology, 2600 GA Delft, The Netherlands}

\author{Arno Bargerbos}
\affiliation{QuTech and Kavli Institute of Nanoscience, Delft University of Technology, 2600 GA Delft, The Netherlands}

\author{Yu Liu}
\affiliation{Center for Quantum Devices, Niels Bohr Institute, University of Copenhagen, 2100 Copenhagen, Denmark}

\author{Leo P. Kouwenhoven}
\affiliation{QuTech and Kavli Institute of Nanoscience, Delft University of Technology, 2600 GA Delft, The Netherlands}

\author{Christian Kraglund Andersen}
\affiliation{QuTech and Kavli Institute of Nanoscience, Delft University of Technology, 2600 GA Delft, The Netherlands}

\date{July 28, 2023}

\maketitle

\tableofcontents

\newpage
 \section{\label{sec:theory} Theoretical description of longitudinal ASQ-ASQ coupling}

\subsection{General description of the estimation of $J$ used in the main text}

We derive a general expression for the coupling strength $J$ in terms of Andreev current operators. The derived expression facilitates the data analysis presented in the main text, where we use the experimentally obtained current-phase relationship, which differs from that expected from the ideal quantum dot junction theory \cite{Padurariu2010, Bargerbos2022b}. The current operator for each individual ASQ can be expressed as $\hat{I_i} = -\frac{2\pi}{\Phi_0}\frac{\partial H_i}{\partial \phi_i}$, where $i=1,2$. Here,  $\Phi_0=h/2e$ denotes the magnetic flux quantum,~$H_i=-\frac{\hbar \omega_i(\phi_i)}{2} \sigma^z_i$ in the subspace of the two spinful doublet states, $\phi_i$ is the phase drop across ASQ$i$, $\sigma_i^z$ is the $z$ Pauli matrix for ASQ$i$ and the $z$ axis is chosen along the spin-polarization direction for each qubit. As a result, the current operator can be related to the qubit frequency by
\begin{equation}\label{eq:current_operator_from_derivative}
\hat{I_i}= \frac{\pi h}{\Phi_0} \frac{\partial f_i(\phi_i)}{\partial \phi_i}\sigma^z_{i} = \frac{I_i}{2}\sigma^z_i,
\end{equation}
where we have defined the amplitude of the spin-dependent current $I_i =\frac{2\pi h}{\Phi_0} \frac{\partial f_i(\phi_i)}{\partial \phi_i} \approx h\frac{\partial f_i(\Phi_i)}{\partial \Phi_i} $ as in the main text, where the last approximation holds in the limit of $L_{\rm J,C} \ll L_{{\rm J},i}^I, L_{{\rm J},i}^\sigma  \forall i$, such that the phase drop can be directly related to the external flux applied through the loop: $\phi_i = \frac{2\pi}{\Phi_0}\Phi_i$. In the subspace of the doublet states for each ASQ we can expand the two-qubit Hamiltonian to first order around the phase bias $\phi_1$, given by the perturbation of the current through ASQ2, $\delta\phi_1 = \frac{2\pi}{\Phi_0} M\hat{I_2}$. Here, $M$ denotes an effective mutual inductance that determines how much phase drops over ASQ1 due to a current in ASQ2. We obtain
\begin{equation} \label{eq:derivation}
\begin{split}
H & = H_1(\phi_1 + \delta\phi_1) + H_2(\phi_2) \\
& = H_1(\phi_1 + \frac{2\pi}{\Phi_0} M\hat{I_2}) + H_2(\phi_2) \\
& \approx H_1(\phi_1) +  \frac{2\pi}{\Phi_0}\frac{\partial H_1(\phi_1)} {\partial\phi_1} M \hat{I_2} + H_2(\phi_2) \\
& =  H_1(\phi_1) + H_2(\phi_2)- M\hat{I_1}\hat{I_2}\\
& =  -\frac{\hbar\omega_1}{2}\sigma_{1}^z  -\frac{\hbar\omega_2}{2}\sigma_{2}^z - \frac{1}{4}M{I_1}{I_2}\sigma^z_{1}\sigma^z_{2}.
\end{split}
\end{equation}
In the limit of $L_{\rm J,C} \ll L_{{\rm J},i}^\sigma \forall i$, where $L_{{\rm J},i}^\sigma$ is the spin-dependent Josephson inductance of ASQ$i$, $M$ is given by the parallel combination of the spin-independent inductances of the three SQUID branches, $M=\frac{L_{\rm J, C} L_{\rm ASQ}}{L_{\rm J, C}+L_{\rm ASQ}}$. Here, $L_{\rm ASQ}(\phi_1, \phi_2)$ is the parallel combination of the spin-independent Josephson inductances of the ASQs:
$$ 
\frac{1}{L_{\rm ASQ}(\phi_1, \phi_2)} = \frac{\cos(\phi_1)}{L_{{\rm J},1}^I} + \frac{\cos(\phi_2)}{L_{{\rm J},2}^I}.
$$
By comparison to Eq.~(1) in the main text, we thus find
\begin{equation}\label{eq:J_current_derivatives}
J = \frac{M}{2h}{I_1}{I_2}= \frac{1}{2h}\frac{L_{\rm J, C} L_{\rm ASQ}}{L_{\rm J, C}+L_{\rm ASQ}} {I_1}{I_2}.
\end{equation}

\subsection{Analytical and numerical calculation of $J$ assuming a sinusoidal current-phase relation}
A simple model of the Hamiltonian for each ASQ is given by~\cite{Padurariu2010, Bargerbos2022b}
\begin{equation}
H_i(\phi_i) = -E^I_{{\rm J}, i} \cos{\phi_i} +  E^\sigma_{{\rm J}, i} \sigma_i^z \sin{\phi_i}, 
\label{Eq:ASQ-Hamiltonian}
\end{equation}
where $E^I_{{\rm J}, i} = \Phi_0^2/(4\pi^2 L^I_{{\rm J}, i})$ and  $E^\sigma_{{\rm J}, i} = \Phi_0^2/(4\pi^2 L^\sigma_{{\rm J}, i})$ denote the spin-independent and spin-dependent Josephson energies, respectively.
The total Hamiltonian of the coupled system of Fig.~1(a) in the main text is thus
\begin{align}
H(\phi) &\,= H_1(\varphi_1-\phi) +  H_2(\varphi_2-\phi) + E_{\rm J, C} \cos{(\phi)} \\
&\,=  -E^I_{{\rm J},1} \cos{(\varphi_1-\phi)} +  E^\sigma_{{\rm J}, 1} \sigma_1^z \sin{(\phi-\varphi_1)}  -E^I_{{\rm J},2} \cos{(\varphi_2-\phi)} +  E^\sigma_{{\rm J}, 2} \sigma_2^z \sin{(\varphi_2-\phi)} + -E_{\rm J, C} \cos{\phi}, 
\label{Eq:total-Hamiltonian}
\end{align}
where $E_{\rm J, C} = \Phi_0^2/(4\pi^2 L_{\rm J, C})$ and the reduced flux, $\varphi_i = 2 \pi \Phi_i /\Phi_0$, is the magnetic flux through the loop containing ASQ$i$ expressed in units of phase.

\subsubsection{Analytical solution}
Following Ref.~\cite{Padurariu2010}, assuming the energy-phase relation in Eq.~\eqref{Eq:ASQ-Hamiltonian}, the lowest order in $E^\sigma_{{\rm J}, 1}/E_{\rm J, C}$ and $E^\sigma_{{\rm J}, 2}/E_{\rm J, C}$ yields the coupling energy in the form
\begin{equation}
J = - 2\frac{E^\sigma_{{\rm J}, 1}E^\sigma_{{\rm J}, 2}}{|\tilde{E}|}   \cos{(\varphi_{\tilde{E}}-\varphi_1)}  \cos{(\varphi_{\tilde{E}}-\varphi_2)}, 
\label{Eq:coupling-Yuli}
\end{equation}
where
\begin{equation}
\tilde{E} = E^I_{\rm J,1}e^{i\varphi_1} + E^I_{\rm J,2}e^{i\varphi_2} + E_{\rm J, C}.
\end{equation}

\subsubsection{Numerical diagonalization}
To go beyond the limit of~\cref{Eq:coupling-Yuli}, i.e. for strong coupling, where the phase-drop on each ASQ is no longer linearly related to the applied flux, we solve the eigenenergies of the system numerically. For a given set of parameters, the energies of the four possible states of the qubit-qubit system ($E_{\uparrow \uparrow}$,  $E_{\downarrow \uparrow}$,  $E_{\uparrow \downarrow}$ and  $E_{\downarrow \downarrow}$) are obtained as the minima in $\phi$ of the four eigenvalues of $H(\phi)$. From these four energies, we calculate the coupling strength $J$ given the longitudinal (or Ising) type coupling Hamiltonian presented in Eq.~(1) in the main text. In this situation, the four eigenenergies of the coupled system are
\begin{align}
E_{\uparrow \uparrow} &\,= \frac{\hbar\omega_1}{2} + \frac{\hbar\omega_2}{2} + \frac{hJ}{2}, \\
E_{\downarrow \uparrow} &\,=  -\frac{\hbar\omega_1}{2} + \frac{\hbar\omega_2}{2} -\frac{hJ}{2},  \\
E_{\uparrow \downarrow} &\,=  \frac{\hbar\omega_1}{2} - \frac{\hbar\omega_2}{2} -\frac{hJ}{2}, \\
E_{\downarrow \downarrow} &\,=  -\frac{\hbar\omega_1}{2}  -\frac{\hbar\omega_2}{2} + \frac{hJ}{2}.
\label{Eq:eigenenergies}
\end{align}
Thus, from the numerically solved eigenenergies, we can find $J$ as 
\begin{equation}
J =  \frac{1}{2h} ( E_{\uparrow \uparrow} - E_{\uparrow \downarrow} - E_{\downarrow \uparrow} + E_{\downarrow \downarrow}).
\label{Eq:numerical_J}
\end{equation}

\subsubsection{Numerics including the transmon degree of freedom}

To fit the transmon spectroscopy data presented in Sec.~\ref{Sss:setpoint} and \ref{s:new-dataset}, we add a charging energy term to~\cref{Eq:total-Hamiltonian} corresponding to the transmon island and numerically diagonalize the resulting Hamiltonian in the phase basis~\cite{Bargerbos2020, Kringhoj2020b}
\begin{equation}
H_\mathrm{Transmon} =  -4E_{\rm c}\partial_\phi^2 + H(\phi)
\label{Eq:total-transmon-Hamiltonian}
\end{equation}
where $E_{\rm c}$ denotes the charging energy of the transmon island and $H(\phi)$ is defined in~\cref{Eq:total-Hamiltonian}.

\subsection{Method comparison}

\begin{figure}[h!]
    \center
    \includegraphics[scale=1.0]{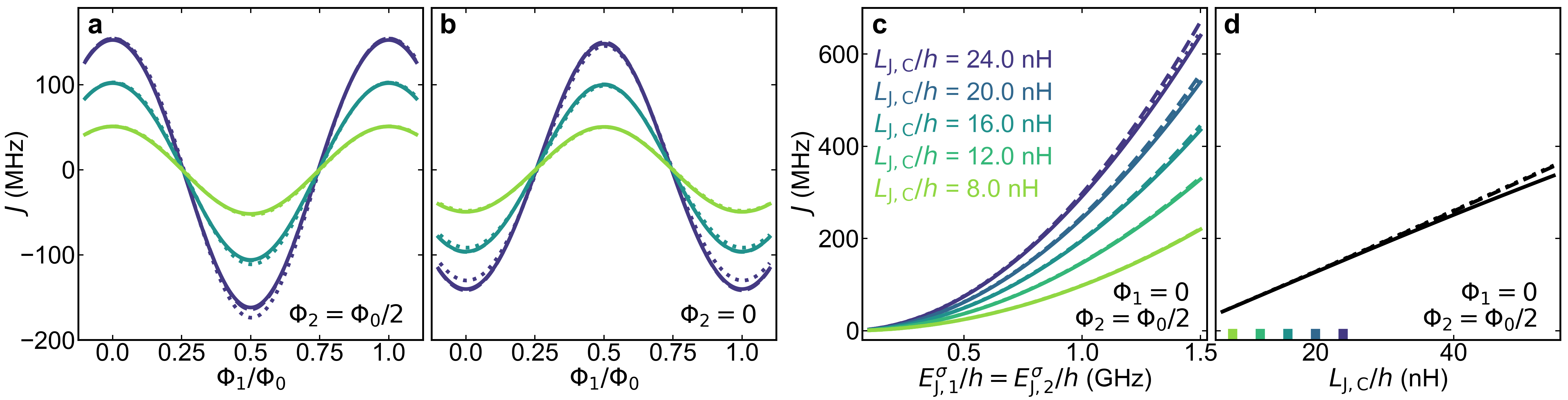}
    \caption{ {\bf Qubit-qubit coupling strength $\mathbf{J}$ as a function of model parameters.} 
    {\bf a}, {\bf b} $\Phi_1$ dependence of the coupling strength $J$ at fixed $\Phi_2 = \Phi_0/2$ and $\Phi_2 = 0$, respectively. 
    {\bf c} $E^\sigma_{{\rm J}, i}$ dependence of $J$ at fixed  $\Phi_1 = 0$ and $\Phi_1 = \Phi_0/2$, for various $L_{\rm J, C}$ values. 
    {\bf d} $L_{\rm J, C}$ dependence of $J$ at fixed  $\Phi_1 = 0$ and $\Phi_1 = \Phi_0/2$. For all panels $E^\sigma_{{\rm J}, 1}/h =$~\SI{0.82}{GHz}, $E^\sigma_{{\rm J}, 2}/h = $~\SI{0.63}{GHz}, $E^I_{\rm J, 1}/h = $~\SI{0.2}{GHz} and $E^I_{\rm J, 2}/h = $~\SI{0.3}{GHz}, excepting for panel c where the values of $E^\sigma_{{\rm J}, 1}$ and $E^\sigma_{{\rm J}, 2}$ are varied. The continuous lines indicate the results obtained from direct diagonalization of Hamiltonian~\eqref{Eq:total-Hamiltonian} using Eq.~\eqref{Eq:numerical_J},  the dashed lines, which mostly fall on top of the solid lines, indicate the analytic limit of Eq.~\eqref{Eq:coupling-Yuli}~\cite{Padurariu2010} and the dotted lines indicate the limit of Eq.~\eqref{eq:J_current_derivatives} used in the main text. Note that, for panel {\bf d}, the values of $I_i\sim h\frac{\partial f_i(\Phi_i)}{\partial \Phi_i}$ are calculated for each value of $L_{\rm J, C}$. This differs from what is done in Fig.~4e in the main text, where the values of $I_i$ are estimated at a fixed $L_{\rm J, C}$ point and used for the complete $L_{\rm J, C}$ range.    }
    \label{fig:supplement-theory} 
\end{figure}

\begin{figure}[h!]
    \center
    \includegraphics[scale=1.0]{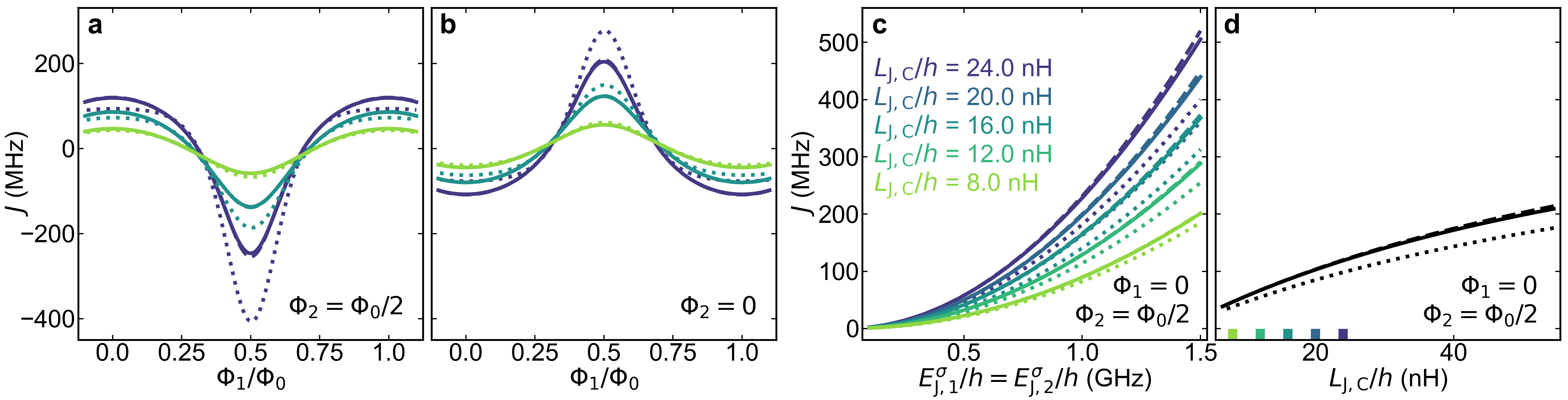}
    \caption{ Same as Fig.~\ref{fig:supplement-theory} but for  $E^I_{\rm J, 1}/h = $~\SI{2.30}{GHz} and $E^I_{\rm J, 2}/h = $~\SI{0.45}{GHz}.
    }
    \label{fig:supplement-theory-realparameters} 
\end{figure}

Given the different approaches to calculate $J$, we now compare the different methods assuming the sinusoidal energy-phase  of~\cref{Eq:ASQ-Hamiltonian}, see Figs.~\ref{fig:supplement-theory} and \ref{fig:supplement-theory-realparameters}.
The analytical expression of~\cref{Eq:coupling-Yuli} is indicated with dashed lines.
The continuous lines are obtained numerically from exact diagonalization of the total Hamiltonian in~\cref{Eq:total-Hamiltonian} using~\cref{Eq:numerical_J}.  
The numerical diagonalization and the analytical expression of~\cref{Eq:coupling-Yuli} show near perfect agreement. 
Only when $E_{J,i}^\sigma \sim E_{J,C}$ (~\cref{fig:supplement-theory}c) a slight deviation is visible since~\cref{Eq:coupling-Yuli} is only valid in the limit $ E_{J,i}^\sigma \ll E_{J,C}$. 
We then test the estimate of $J$ on the sinusoidal energy-phase relation of~\cref{Eq:ASQ-Hamiltonian} using~\cref{eq:J_current_derivatives}, which is also used in the main text for the experimentally obtained energy-phase relation. This is shown with dotted lines, for different sets of parameters. 
In Fig.~\ref{fig:supplement-theory} we use parameters corresponding to the limit $L_{\rm J,C} \ll L_{{\rm J},i}^\sigma, L_{{\rm J},i}^I\forall i$ and, given the agreement between the different methods, we note that the approximations made in Sec.~\ref{sec:theory} are valid. Thus, the general estimate from Eq.~\eqref{eq:J_current_derivatives} (dotted line) agrees well with the exact value of $J$ found by numerical diagonalization of the full Hamiltonian, as expected. 
To illustrate the estimates obtained from the different methods outside of this limit, we use values of $E^I_{{\rm J}, i}$ in Fig.~\ref{fig:supplement-theory-realparameters}
that instead deviate from the limit $L_{\rm J,C} \ll  L_{{\rm J},i}^I\forall i$. In this case, we see that the estimate from Eq.~\eqref{eq:J_current_derivatives} deviates strongly from the exact numerical calculation due to the non-linear flux-phase relation.

\FloatBarrier

\subsection{Master equation approach to longitudinal coupling experiment}
\begin{figure*}[h]
\includegraphics{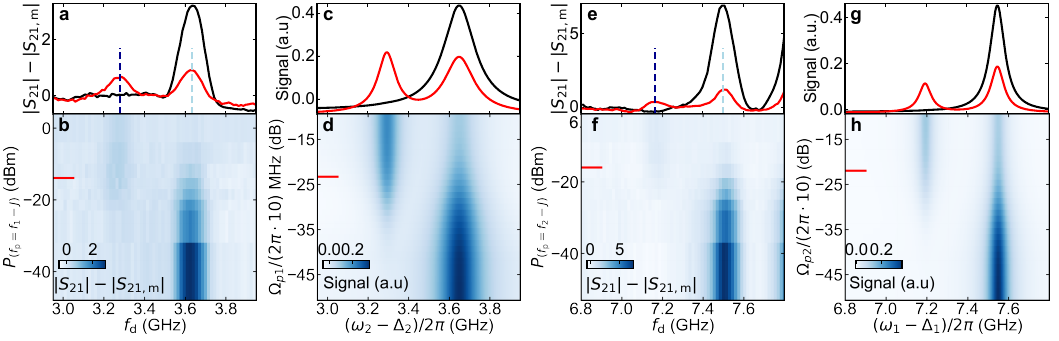}
\caption{{\bf Comparison between experiment and master-equation simulation.}
{\bf a}-{\bf b}, {\bf e}-{\bf f} Experimental data of longitudinal coupling measurement repeated from Fig.~3 of the main text.
{\bf c}-{\bf d}, {\bf g}-{\bf h} Results of master equation simulations of the corresponding experimental data.
{\bf c}-{\bf d} ASQ1 is driven with a pump tone at $f_{\rm p} = f_1-J$, while doing spectroscopy on ASQ2.
{\bf g}-{\rm h} ASQ2 is driven with a pump tone at $f_{\rm p} = f_2-J$, while doing spectroscopy on ASQ1.
We use the following parameters: the spectroscopy drive amplitude for ASQ$i$ is set to $\Omega_{pi}/2\pi=\SI{2}{\mega\hertz}$ which power broadens the observed linewidths similar to the experiment. The drive frequency of the third tone is set such that $\Delta_i=-J$, and the power is shown on the y-axis of the 2D maps in dB, similar to the experiment. $T_1$ and  $T_2^*$ are set to their values shown in Fig.~2 of the main text using the collapse operators and $\omega_i/(2\pi)$ of ASQ$i$ are set to $f_i-J$. The dispersive shifts are assumed larger than the linewidth of the resonator here such that the signal is only sensitive to the change in $\ket{\downarrow_1\downarrow_2}$ population. In all simulations, we fix $J=\SI{178}{\mega\hertz}$.}
\label{fig:supplement-qutip-fig3}
\end{figure*}

We now present a simple master equation simulation to investigate the effect of the drives on the coupled two-qubit system in presence of decay. We solve the Lindblad master equation for the time evolution of the system density matrix, $\rho$, of the following form 
\begin{equation}\label{eq:lindblad_me}
\dot{\rho} = \left[\rho, H'\right] - \sum_n\frac{1}{2}\left[2C_n\rho C_n^\dagger - \rho C_n^\dagger C_n - C_n^\dagger C_n \rho\right],
\end{equation}
where $H'$ describes the two-qubit system in the rotating frame of the two drives, which have certain detuning $\Delta_i$ from qubit $i$.
This results in the following Hamiltonian
\begin{equation}
H'/\hbar = \frac{\Delta_1}{2}\sigma^z_{2} + \frac{\Delta_2}{2} \sigma^z_{2} + \frac{\Omega_{p1}}{2} \sigma^x_{1} + \frac{\Omega_{p2}}{2}\sigma^x_{2} + 2\pi \frac{J}{2}\sigma^z_{1}\sigma^z_{2},
\end{equation}
where $\Omega_{pi}$ denotes the drive amplitude of the tone near qubit $i$ and $\Delta_i = \omega_i - \omega_{pi}$ is the detuning of that drive frequency with the qubit frequency. 
Additionally, we apply the collapse operators $C_n$ on the individual qubits to simulate the effect of finite $T_1$ and $T_2$: $C_n \in \{\sqrt{\gamma_{1,1}}\sigma_1^-, \sqrt{\gamma_{\phi_1}/2} \sigma^z_{1}\, \sqrt{\gamma_{1,2}}\sigma_2^-, \sqrt{\gamma_{\phi_2}/2} \sigma^z_{2}\}$, where $\gamma_{1, i}=1/T_1^{{\rm ASQ}i}$,  $\gamma_{\phi_i}=1/T_2^{{\rm ASQ}i}$, $\sigma_i^+ = \ket{\uparrow_i}\bra{\downarrow_i}$ and $\sigma_i^- = \ket{\downarrow_i}\bra{\uparrow_i}$. We then solve~\cref{eq:lindblad_me} for the steady state solution using Qutip~\cite{qutip2013}. 
From the above evolution of the master equation under a certain drive amplitude, we obtain the populations of the states $\left\{\ket{\uparrow_1\uparrow_2}, \ket{\uparrow_1\downarrow_2},\ket{\downarrow_1,\uparrow_2},\ket{\downarrow_1\downarrow_2}\right\}$. Then, assuming a dispersive shift for each state and a linewidth of the resonator mode, we calculate the signal as the sum of populations times the displaced Lorentzians corresponding to each state and subtract the median for each linecut as is done with the experimental data.   
We compare the data measured in Fig.~3 of the main text to the master equation simulation with realistic parameters, as shown in~\cref{fig:supplement-qutip-fig3}. The simulations reproduce the main features seen in the data. 

The peak height difference between the drive being on and off (black and red linecuts in~\cref{fig:supplement-qutip-fig3}) depends on the difference between the initial and final populations $P_{\downarrow\downarrow}$ of $\ket{\downarrow_1\downarrow_2}$ in the limit of large dispersive shift, which we consider here for simplicity. Consider the case where we apply a spectroscopy tone at $f_2-J$ on ASQ2, in the absence of a pump tone. In steady state, we get $P_{\downarrow\downarrow} =P_{\downarrow\uparrow}=0.5$ and $P_{\uparrow\uparrow} = P_{\uparrow\downarrow}=0$ 
due to the spectroscopy saturating ASQ2. Now, if we set a separate pump tone driving ASQ1 at $f_{\rm p}=f_1-J$ to a sufficiently high amplitude $\Omega_{p1}$, we obtain $P_{\downarrow\downarrow}= P_{\downarrow\uparrow} = P_{\uparrow\downarrow} = 0.33$. 
Thus the height of the driven peak at $f_2+J$ (red right peak) should be the height of the undriven (black) peak divided by a factor of $0.5/(0.5-0.33)\sim 2.94$ (as opposed to a factor of 2 which one might naively expect). The residual lowering of the peak observed in the experiment, we attribute to additional losses in the resonator mode under a strong drive. The height of the peak at $f_2-J$, on the other hand, is expected to have a similar height if no $T_1$ decay is present. 
In presence of finite and similar $T_1$ for ASQ1 and ASQ2, however, the final populations end up becoming $P_{\downarrow\downarrow}= P_{\downarrow\uparrow} = P_{\uparrow\downarrow} = P_{\uparrow\uparrow} = 0.25$, thus increasing the signal at $f_2-J$ and leading to a higher peak reaching half the height of the undriven peak at $f_2+J$ (as also seen in in~\cref{fig:supplement-qutip-fig3}(d)). However, beyond these limiting cases, depending on the exact ratio of the $T_1$ lifetimes of ASQ1 and ASQ2 the steady-state populations will vary.

\newpage
\section{\label{Ss:device} Methods}

\subsection{Device overview}

\begin{figure}[p]
    \center
    \includegraphics[scale=1.0]{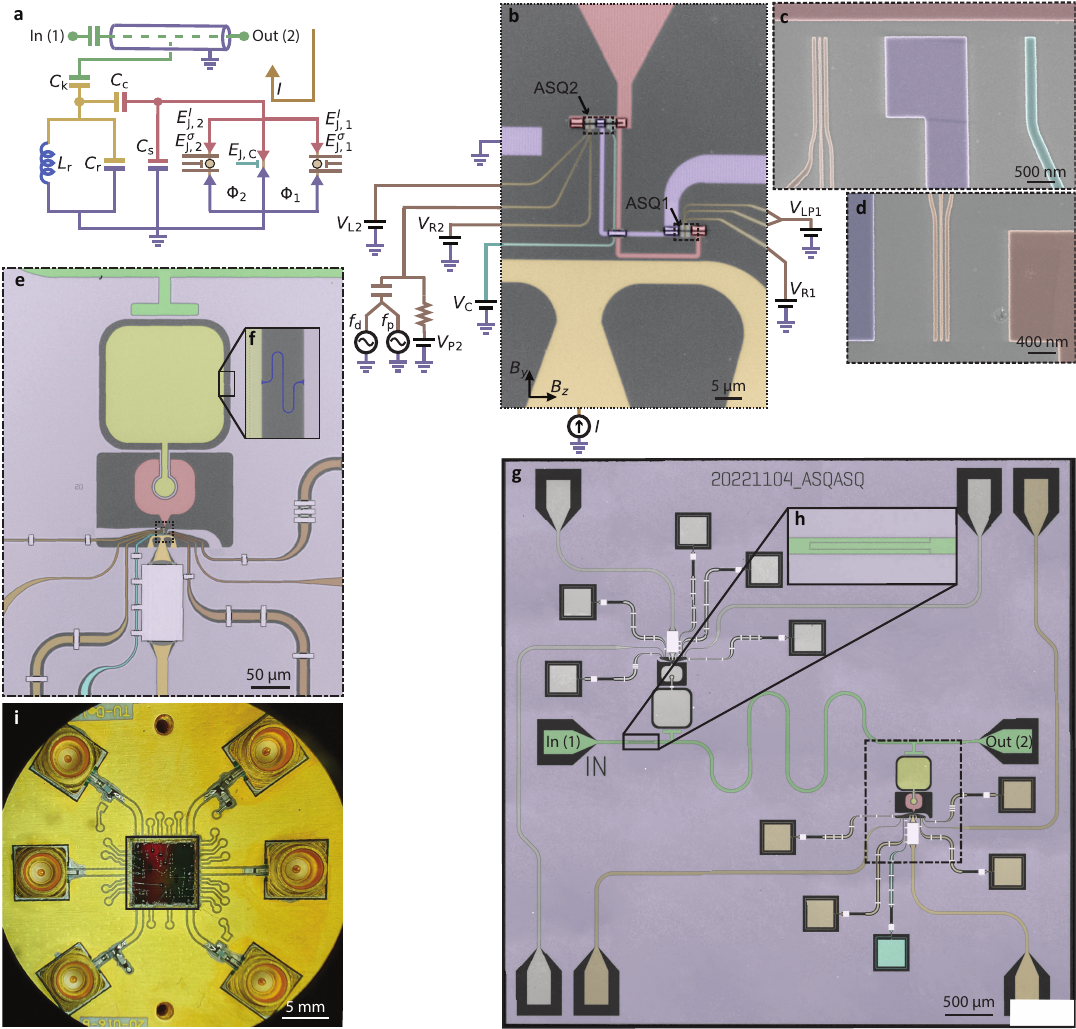}
    \caption{{ \bf Device overview.}  {\bf a} Diagram of the full microwave circuit. A coplanar waveguide (green center conductor) transmission line with an input capacitor is capacitively coupled to a grounded LC resonator. The resonator consists of an island (yellow) capacitively and inductively  (blue) shunted to ground (purple). The resonator is in turn capacitively coupled to a transmon island (red), which is shunted to ground capacitively as well as via three parallel Josephson junctions. The coupling junction is controlled by a single electrostatic gate (cyan) and each of the two Andreev spin qubits is controlled by three electrostatic gates (brown). The RF drive tones $f_{\rm d}$ and $f_{\rm p}$ are sent through the plunger gate of ASQ2. The current through the flux line (amber), $I$,  controls the flux thread through the loop containing ASQ1, $\Phi_1$, and leaves $\Phi_{\rm 2}$ nearly unaffected.
    {\bf b} False-colored optical microscope image of the loops area. The three Josepshon junctions are implemented in two separate Al/InAs nanowires, one of them containing the coupling junction and ASQ2 and the other containing ASQ1.  The $B_y$ component of the magnetic field is used to tune $\Phi_1$ and $\Phi_2$, see ~\cref{fig:loops_diagram} for detailed sketch of the loops geometry. $B_z$ is the magnetic field component approximately parallel to the nanowires axis. 
    {\bf c}, {\bf d} False colored scanning electron microscope (SEM) images of the gates areas taken before the deposition of the gate dielectric and nanowire.
    {\bf e} False-colored optical microscope image of the device showing the qubit island (red), the resonator island (yellow), the \SI{200}{nm}-wide resonator inductor (blue, enlarged in {\bf f}), the transmission line (green), the electrostatic gates (brown and cyan) the flux line (amber) and ground (purple). 
    {\bf g}  False-colored optical microscope image of the whole 6$\times$6~mm chip containing two nearly identical devices coupled to the same transmission line, which has an input capacitor, enlarged in {\bf h}.  The measured device is false-colored, while the second device was not investigated.
    {\bf i}  Chip mounted on a printed circuit board (PCB).
    }
    \label{fig:device} 
\end{figure}

The physical implementation of the device investigated is shown in Fig.~\ref{fig:device}. 
The chip, 6~mm long and 6~mm wide, consists of two devices coupled to a single transmission line with an input capacitor to increase the directionality of the outgoing signal (Fig.~\ref{fig:device}h).
For the experiments performed here, only the device discussed in the main text, highlighted in Fig.~\ref{fig:device}g, was measured. The resonator of the second device (uncolored device in Fig.~\ref{fig:device}g) was not functional and thus was not investigated.

For each device, a lumped element readout resonator is capacitively coupled to the feedline (Fig.~\ref{fig:device}e). The resonator is additionally capacitively coupled to the transmon island, which is connected to ground via  three Josephson junctions in parallel (the coupling junction, ASQ1 and ASQ2) defining two loops (Fig.~\ref{fig:device}b). The three junctions are implemented on two separate Al/InAs nanowires. 
The junctions are defined by etching the aluminum shell of the nanowire in a \SI{95}{nm}-long section for the coupling junction and \SI{215}{nm}-long sections for each of the ASQ junctions.  
The coupling junction is controlled by a single \SI{200}{nm}-wide electrostatic gate centered at the middle of the junction, controlled with a DC voltage \Vc. 
Each of the quantum dot junctions is defined by three gates consisting of two \SI{50}{nm} wide tunnel gates (L, R) surrounding a \SI{60}{nm} wide plunger gate (P), separated from each other by \SI{45}{nm} (Fig.~\ref{fig:device}c, d).
We define the DC voltages used for the left and right tunnel gates as \Vlpa~and \Vra~for ASQ1 or \Vlb~and \Vrb~for ASQ2. The plunger gate of ASQ1 is also set to \Vlpa because it was shorted to the left tunnel gate due to a fabrication imperfection. 
All gate lines except for the plunger lines incorporate a fourth-order Chebyshev LC-LC filter with a cut-off frequency at \SI{2}{GHz}. The first and second inductive elements, of \SI{5.2}{nH} and \SI{6.1}{nH} respectively, are implemented using thin strips of NbTiN with widths of \SI{3.5}{\micro m} and \SI{300}{nm}, respectively. The first and second capacitive elements, of \SI{2.45}{pF} and \SI{2.08}{pF} respectively, are implemented with parallel plate capacitors.
The plunger gate of ASQ2  is connected to a bias-tee on the printed circuit board formed by a \SI{100}{\kilo\ohm} resistor and a \SI{100}{pF} capacitor. This permits the simultaneous application of a DC signal, \Vpb, to control the level of the quantum dot junction, and microwave tones, $f_{\rm d}$ and $f_{\rm p}$, to drive either of the spin-flip transitions or the transmon. We also drive ASQ1 using the same gate line, because the bias-tee at the plunger gate of ASQ1 was not functional. The flux through the loop containing ASQ1 is controlled using a flux line (shown in amber). Its design in the area of the loops was inspired by Ref.~\cite{Rot2022}. We furthermore incorporate a  \SI{25}{pF}  parallel plate capacitor near the end of the flux line which, together with the \SI{1}{nH} inductance of the rest of the flux line, implements an LC low-pass filter with a cut-off at \SI{1}{GHz}.

\begin{figure}[h!]
    \center
    \includegraphics[scale=0.9]{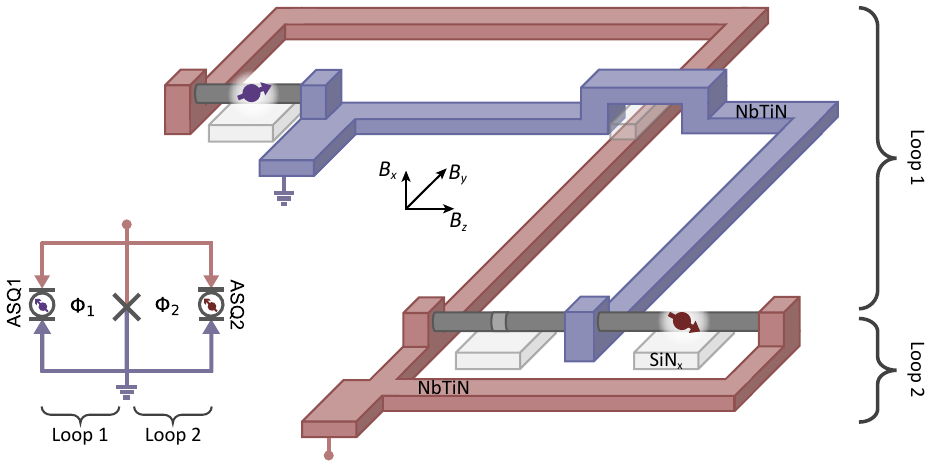}
    \caption{{ \bf Loops geometry.} 
    Diagram of the loops area shown in Fig.~\ref{fig:device}b.
    Red and purple NbTiN segments denote segments connected to the transmon island and ground, respectively.
    The loop containing the coupling junction and ASQ2 (loop 2) is a planar loop with the same geometry as those in Refs.~\cite{Bargerbos2022, Bargerbos2022b, PitaVidal2023}.
    The loop containing the coupling junction and ASQ1 (loop 1) is a twisted gradiometric loop formed by two subloops. Equal of out-of-plane magnetic fields $B_x$ through each of the two subloops result in nearly opposite contributions to the flux $\Phi_1$, rendering loop 1 nearly insensitive to out-of-plane magnetic field noise.
    The nanowires are elevated with respect to the NbTiN plane due to the gate dielectric (light grey). This defines, for each loop, a loop area perpendicular to $B_y$. $B_y$ can thus be used to control the flux through the loops while keeping the out-of-plane field component ($B_x$) fixed, reducing the occurrence of external flux jumps~\cite{Wesdorp2022}.
    }
    \label{fig:loops_diagram} 
\end{figure}

\subsection{Summary of device parameters}\label{Sss:parameters_table}

\begin{table*}[h!]
\begin{ruledtabular}
\begin{tabular}{rcrc}
Bare resonator frequency, $f_{\rm r, 0}$ & \SI{4.229}{GHz} & Resonator $Q_{\rm c}$ & 1.3k \\
Resonator $Q_{\rm i}$ & $\sim$~35k & Transmon decay time, $T_{1}^{\rm t}$ & \SI{53.6}{ns}  \\
Resonator-transmon coupling, $g/h$ & $\sim$~\SI{287}{MHz}   & Transmon Ramsey time, $T_{\rm 2R}^{\rm t}$ & \SI{80.0}{ns}   \\
Transmon charging energy, $E_{\rm c}/h$ &   \SI{200}{MHz} &
\\
\end{tabular}
\end{ruledtabular}
\caption{
{\bf Values of relevant device parameters.} The resonator bare frequency and quality factors are measured when all electrostatic gates are at \SI{-1000}{mV} and thus all three junctions are pinched off (see Fig.~\ref{fig:resonator}). The transmon charging energy is extracted from the transmon anharmonicity in two-tone spectroscopy. The resonator-transmon coupling is extracted from a single-tone spectroscopy measurement at their anti-crossing (see Fig.~\ref{fig:supplement-pinchoff}). The transmon coherence values were measured with both ASQs in pinch off and at $V_{\rm C}$~=~\SI{1500}{mV}, which sets the transmon frequency to $f_{\rm t}$~=~\SI{5.45}{GHz}. 
}
\label{tab:circuit_parameters}
\end{table*}

\subsection{Nanofabrication details}

The device fabrication occurs in several steps identical to that described in~\cite{Bargerbos2022}, and repeated here for the sake of completeness. The substrate consists of 525~$\upmu$m-thick high-resistivity silicon, covered in \SI{100}{nm} of low-pressure chemical vapor deposited $\rm{Si_3N_4}$. In the first step, a 4-inch wafer of such substrate is cleaned by submerging it for \SI{5}{min} in HNO$_3$ while ultasonicating, followed by two short H$_2$O immersions to rinse the HNO$_3$ residues. Afterwards, a \SI{20}{nm}-thick NbTiN film is sputtered on top of the substrate using an \textit{AJA International ATC 1800} sputtering system. Subsequently, Ti/Pd e-beam alignment markers are patterned on the wafer, which is thereafter diced into smaller individual dies of approximately \SI{12}{mm}$\times$\SI{12}{mm}. In the next step, the gate electrodes and the rest of the NbTiN circuit elements are patterned on one die covered by \SI{110}{nm}-thick \textit{AR-P 6200} (positive) e-beam resist using electron-beam lithography. The structures are then etched using $\rm{SF_6}$/$\rm{O_2}$ reactive ion etching for \SI{47}{s}. Subsequently, \SI{28}{nm} of $\rm{Si_3N_4}$ dielectric are deposited on top of the gate electrodes using plasma-enhanced chemical vapor deposition and etched in patterns with a buffered oxide etchant (for \SI{3}{min}). This dielectric is used as a gate dielectric, as well as as the dielectric for the crossovers at the DC gate lines and flux line and for the crossover that generates the twist in the loop containing ASQ1. 

The nanowires are deterministically placed on top of the dielectric using a nanomanipulator and an optical microscope.  These nanowires are $\sim$\SI{10}{\micro m}-long epitaxial superconductor-semiconductor nanowires with a \SI{110}{nm}-wide hexagonal InAs core and a \SI{6}{nm}-thick Al shell covering two of their facets, in turn covered by a thin layer of aluminium oxide. The growth conditions were almost identical to those detailed in Ref.~\cite{Krogstrup2015}, with the only two differences being that this time the As/In ratio was 12, smaller than in Ref.~\cite{Krogstrup2015}, and that the oxidation of the Al shell was now performed in-situ, for better control, reproducibility and homogeneity of the oxide layer covering the shell. Inspection of the nanowire batch, performed under a scanning electron microscope directly after growth, indicated an average wire length of $9.93 \pm$~\SI{0.92}{\micro m} and an average wire diameter of $111 \pm 5$~nm.

After nanowire placement, three sections of the aluminium shells are selectively removed by wet etching for \SI{55}{s} with \textit{MF-321} developer. 
These sections form the two quantum dot junctions and the coupling junction, with lengths \SI{215}{nm} and \SI{95}{nm}, respectively. After the junctions etch, the nanowires are contacted to the transmon island and to ground by a \SI{110}{s} argon milling step followed by the deposition of \SI{150}{nm}-thick sputtered NbTiN. Finally, the chip is diced into 6 by 6 millimeters, glued onto a solid gold-plated copper block with silver epoxy, and connected to a custom-made printed circuit board using aluminium wire-bonds (Fig.~\ref{fig:device}g).

\subsection{Cryogenic and room temperature measurement setup}

\begin{figure}[h!]
    \center
    \includegraphics[scale=0.58]{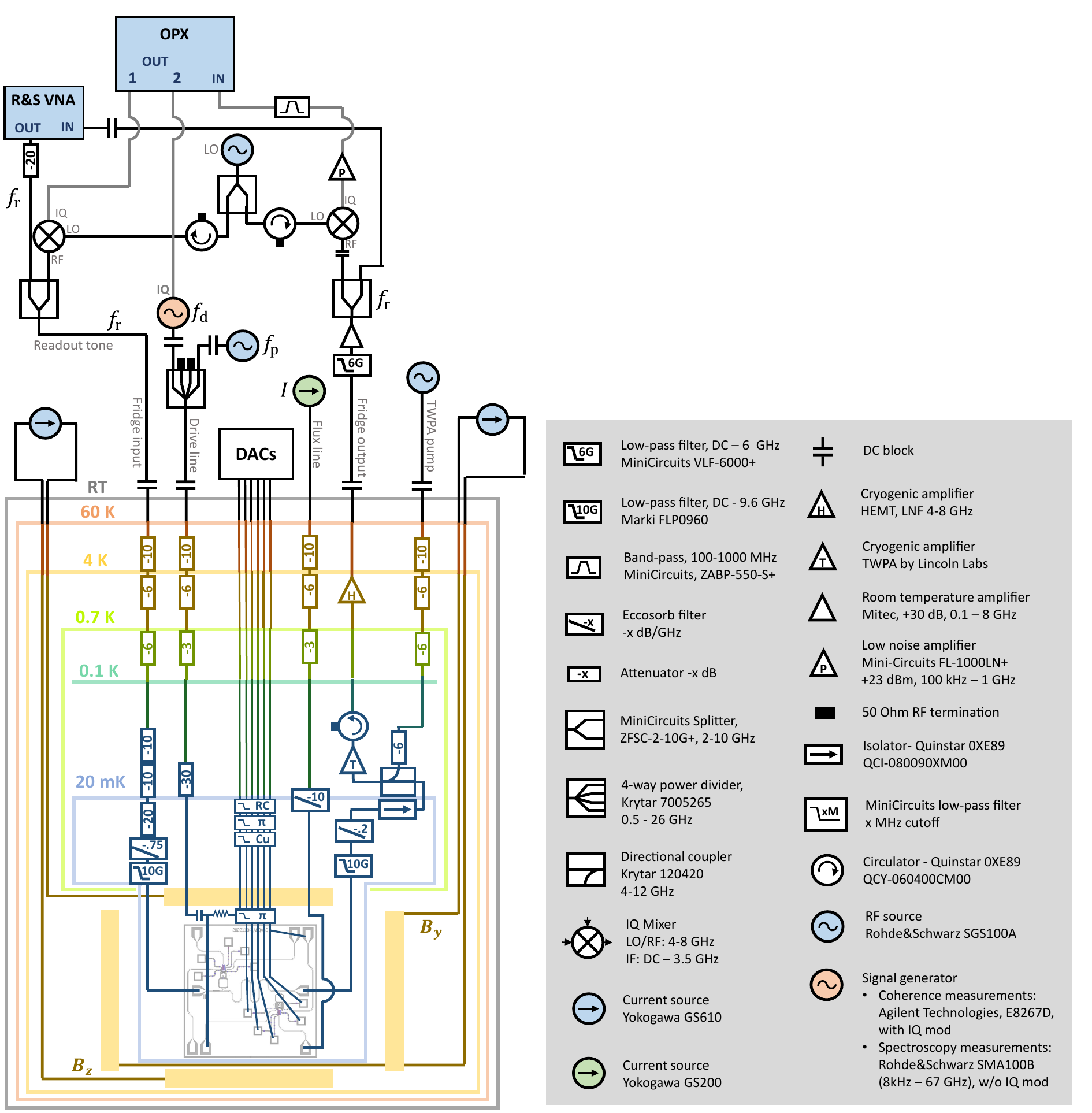}
    \caption{{ \bf Measurement setup.}}
    \label{fig:cryogenic_setup}
\end{figure}

The device was measured in an \textit{Oxford instruments Triton} dilution refrigerator with a base temperature of approximately \SI{20}{mK}. Details of the wiring at room and cryogenic temperatures are shown in Fig.~\ref{fig:cryogenic_setup}. The setup contains an input radio-frequency (RF) line, an output RF line, an extra RF line for the drive tones, a flux-bias line and multiple direct current (DC) lines used to tune the electrostatic gate voltages. The DC gate lines are filtered at base temperature with multiple low-pass filters connected in series. 
The input, flux and drive RF lines contain attenuators and low-pass filters at different temperature stages, as indicated. In turn, the output RF line contains amplifiers at different temperature stages: a traveling wave parametric amplifier (TWPA) at the mixing chamber plate ($\approx$~\SI{20}{mK}),  a high-electron-mobility transistor (HEMT) amplifier at the \SI{4}{K} stage, and an additional amplifier at room temperature. 
A three-axis vector magnet, for which the $y$ and $z$ coils are illustrated by yellow rectangles in Fig.~\ref{fig:cryogenic_setup} ($x$-axis not shown), is thermally anchored to the \SI{4}{K} temperature stage, with the device under study mounted at its center. The three magnet coils are controlled with \textit{Yokogawa GS610} current sources. The current through the flux line, $I$, is controlled with a \textit{Yokogawa GS200} current source.
At room temperature, a vector network analyzer (VNA) is connected to the input and output RF lines for spectroscopy at frequency $f_{\rm r}$. On the input line, this signal is combined with a separate IQ-modulated tone also at $f_{\rm r}$, only used for time-domain measurements. The IQ-modulated drive tone at frequency $f_{\rm d}$ and the pump tone at frequency $f_{\rm p}$ are both sent through the drive line. For time-domain measurements, the output signal is additionally split off into a separate branch and down-converted to be measured with a \textit{Quantum Machines OPX}.

\subsection{\label{Ss:methods} Data processing}

\subsubsection{Background subtraction for single-tone and two-tone spectroscopy measurements}
For all single-tone spectroscopy measurements shown in the main text and Supplementary Information, we plot the amplitude of the transmitted signal, $|S_{21}|$, with the frequency-dependent background, $|{S_{21, {\rm b}}(f_{\rm r})}|$, divided out, in dB:  $10\log_{10}(|S_{21}|/|{S_{21, {\rm b}}}|)$. The background is extracted from an independent measurement of the transmission through the feedline as a function of \Vc. To determine the background for each $f_{\rm r}$ we do not consider transmission data for which the resonator frequency is more than \SI{20}{MHz} close to $f_{\rm r}$, so that the presence of the resonator does not impact the extracted background.

For two-tone spectroscopy measurements, we instead plot the transmitted signal, $|S_{21}|$, with the frequency-independent background, $|{S_{21, {\rm m}}}|$, subtracted: $|S_{21}|-|{S_{21, {\rm m}}}|$. In this case, the background is defined as the median of $|S_{21}|$ of each frequency trace.

\subsubsection{Gaussian fits to extract $J$ from spectroscopy measurements}
To extract the value of the coupling strength from the peak splitting observed in spectroscopy measurements, we follow the following procedure:
\begin{enumerate}
    \item We first fit the two-tone spectroscopy signal in the absence of a pump tone with a single Gaussian function of the form
    \begin{equation}
        \label{eq:single_Gaussian}
        \frac{A}{\sqrt{2\pi\sigma^2}}\exp\left(\frac{-(f_{\rm d}-f_a)^2}{2\sigma^2}\right)+Bf_{\rm d}+C,
    \end{equation}
    from which we extract the position of the first peak, $f_a$, and its width, $\sigma$.
    \item Next, we fit the two-tone spectroscopy signal in the presence of a pump tone with a double Gaussian function of the form
    \begin{equation}
        \label{eq:double_Gaussian}
        \frac{A_a}{\sqrt{2\pi\sigma^2}}\exp\left(\frac{-(f_{\rm d}-f_a)^2}{2\sigma^2}\right)+\frac{A_b}{\sqrt{2\pi\sigma^2}}\exp\left(\frac{-(f_{\rm d}-f_b)^2}{2\sigma^2}\right)+Bf_{\rm d}+C,
    \end{equation}
    for which the peak widths $\sigma$, as well as the position of the first peak, $f_a$, are fixed to their values extracted from the previous fit. From this fit, we extract the position of the second peak, $f_b$, as well as the chi-square of the fit, $\chi^2_{\rm double}$.
    \item Next, we repeat a fit to the two-tone spectroscopy signal in the presence of a pump tone with a single Gaussian function (Eq.~\eqref{eq:single_Gaussian}) and extract the chi-square of the fit, $\chi^2_{\rm single}$.
    \item To determine whether a double Gaussian fits better than a single Gaussian, we compare the goodness of fit of a double and single Gaussian fit. If $(\chi^2_{\rm single}-\chi^2_{\rm double})/\chi^2_{\rm double} \geq 0.1$, we conclude that the data shows two peaks and extract $J$ as $J=(f_a-f_b)/2$ and the error of $J$ as the error of $f_b$ extracted from the double Gaussian fit (its one-sigma confidence interval).
    \item If, else, $(\chi^2_{\rm single}-\chi^2_{\rm double})/\chi^2_{\rm double} < 0.1$, we conclude that the data shows a single peak and thus $J=0$.
\end{enumerate}
    
\subsubsection{Determination of the flux axis}\label{sss:fits_spin_flip}
To determine the flux axis for data that we display in the main text as a function of $\Phi_i$, we map the corresponding flux control parameter ($I$ for the loop containing ASQ1 and $B_y$ for the loop containing ASQ2) to the fluxes $\Phi_1$ and $\Phi_2$. To do so, we need to determine the value of the control parameter corresponding to $\Phi_i=0$ (denoted as $I_{\Phi_1=0}$ and $B_{y, \Phi_2=0}$, respectively) as well as the one flux quanta (denoted as $I_{\Phi_0}$ and $B_{y, \Phi_0}$, respectively). The former is independently determined for each separate measurement, from fits of the data to the expected transitions.

The values of the flux quanta ($I_{\Phi_0}=$~\SI{9.62}{mA} for $\Phi_1$ and $B_{y, \Phi_0}=$~\SI{3.16}{mT} for $\Phi_2$) are fixed throughout all main text and supplementary figures and is extracted from fits to the data in Fig.~2a and b in the main text. The data in Fig.~2b is fitted with a sinusoidal dependence of the form
\begin{equation}
    \label{eq:sin}
    2E^\sigma_{\rm J, 2}\sin\left(2\pi\frac{B_y-B_{y, \Phi_2=0}}{B_{y, \Phi_0}}\right)+C,
\end{equation}
as expected for a quantum dot Josephson junction \cite{Padurariu2010, Bargerbos2022b}. The data in Fig.~2a is instead fitted with a phenomenological skewed sinusoidal dependence of the form
\begin{equation}
    \label{eq:sin_sk}
    2E^\sigma_{\rm J, 1}\sin\left(2\pi\frac{I-I_{\Phi_1=0}}{I_{\Phi_0}} + S\sin\left(2\pi\frac{I-I_{\Phi_1=0}}{I_{\Phi_0}}\right) \right)+C,
\end{equation}
where $-1< S <1$ is the skewness parameter. For the data in Fig.~1a, we extract $S=-0.39$. The observed skewness of the spin-flip is, to our knowledge, not predicted by existing models~\cite{Padurariu2010, Bargerbos2022b}, so further investigation is needed to explain its origin.

\newpage

\section{\label{Ss:tuneup} Basic characterization and tuneup}

\subsection{Readout resonator characterization} \label{Sss:resonator}

\begin{figure*}[h!]
\includegraphics{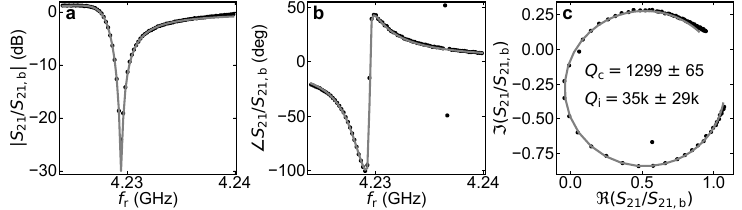}
\caption{{\bf Single-tone spectroscopy of the resonator and quality factor extraction.} All panels show the measured data (black markers) and a fit to Eq.~\eqref{Eq:resonator_fit} (grey line). {\bf a} and {\bf b} show, respectively, the amplitude and phase of the $S_{21}$ signal as a function of frequency. {\bf c} shows the imaginary and real parts of the complex $S_{21}$ signal. From the fit, we extract a resonator bare frequency of $f_{\rm r, 0}=$~\SI{4.22850}{GHz}~$\pm$~\SI{91}{kHz} and the quality factors indicated in {\bf c}.
}
\label{fig:resonator}
\end{figure*}

In this section, we perform a fit to a bare resonator spectroscopy trace and extract the resonator parameters shown in Tab.~\ref{tab:circuit_parameters}.  The result of a single-tone resonator trace, performed with all Josephson junctions pinched off, is shown with black markers in Fig.~\ref{fig:resonator}. The grey lines show the best fit of the complex transmission to the expected dependence \cite{Khalil2012, ResonatorRepo}
\begin{equation}
S_{21}(f_{\rm r}) = 1 - \frac{1 +  i \alpha} { 1 + \frac{Q_{\rm c}}{Q_{\rm i}} + 2  Q_{\rm c} i \frac{f_{\rm r}-f_{\rm r,0}}{f_{\rm r,0}}}, 
\label{Eq:resonator_fit}
\end{equation}
where $f_{\rm r,0}$ is the bare resonator resonance frequency, $Q_{\rm c}$ and $Q_{\rm i}$ are the coupling and internal quality factors, respectively, and $\alpha$ is a real number to account for the resonator asymmetry. 

\subsection{Gate and flux characterization} \label{Sss:parameter}

\begin{figure}[h!]
    \center
    \includegraphics[scale=1.0]{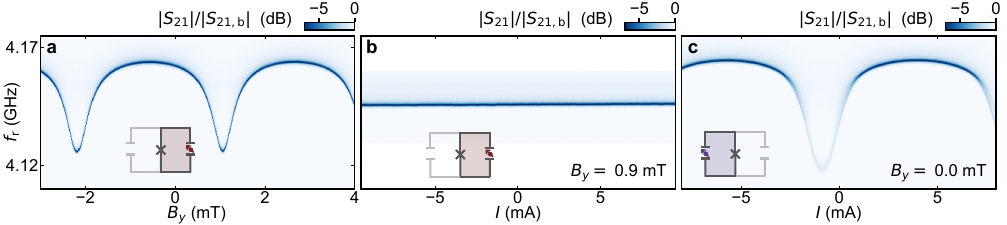}
    \caption{ {\bf Flux control.} 
    {Amplitude of the transmission through the readout circuit, $|S_{21}|$, with background, $|S_{21, \rm b}|$, divided out. }
    {\bf a}, {\bf b} With ASQ1 closed (\Vlpa~=~\Vra~=~\SI{-1000}{mV}) and ASQ2 open to a singlet state with large Josephson energy. {\bf a} shows the dependence on $B_y$, which tunes  $\Phi_{2}$, while {\bf b} shows the dependence on the current through the flux line, $I$, which leaves $\Phi_2$ unaffected.
    {\bf c} With ASQ1 open to a singlet state with large Josephson energy and ASQ2 closed (\Vlb~=~\Vpb~=~\Vrb~=~\SI{-1000}{mV}), plotted versus $I$, which controls $\Phi_1$.
    For all panels, $B_x = B_z = 0$ and \Vc~=~\SI{1995}{mV}.
    }
    \label{fig:flux} 
\end{figure}

Throughout this manuscript, we use $B_y$, which affects both $\Phi_1$ and $\Phi_2$, to tune $\Phi_2$ and the current through the flux line, $I$, to tune $\Phi_1$. Fig.~\ref{fig:flux}a shows the $B_y$ tunability of $\Phi_2$, for which the period corresponds to \SI{3.25}{mT}. Note that this value is slightly larger than the actual flux quantum due to the small flux jumps present in the signal. Fig.~\ref{fig:flux}c and b show how the current through the flux line, $I$, controls $\Phi_1$, for which a flux quantum corresponds to \SI{9.61}{mA}, while leaving $\Phi_2$ unaffected.

\begin{figure*}[h!]
\includegraphics{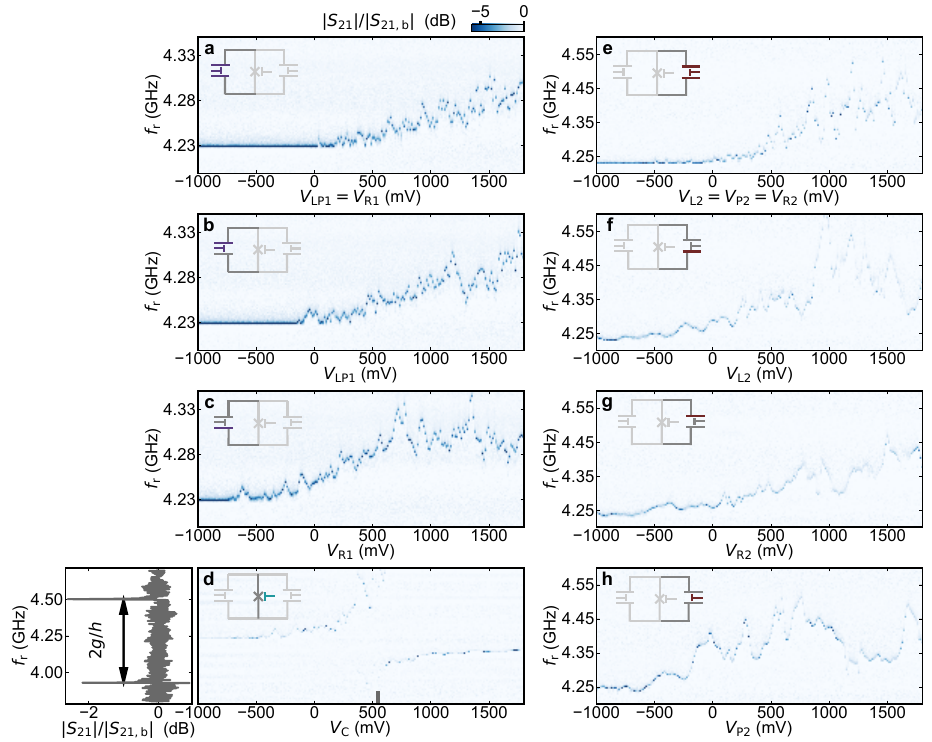}
\caption{{\bf Individual characterization of electrostatic gates.} All panels show single-tone spectroscopy of the resonator versus different gate voltages. The gate voltages that are being varied in each case are highlighted with colors in the insets. Dark grey shaded gates indicate open gates set to \SI{1800}{\milli\volt}. Light grey shaded gates indicate closed gates set to \SI{-1000}{\milli\volt}. The panel to the left of {\bf d} shows a line cut of the data in {\bf d} taken at the \Vc~value indicated with a grey line in the x-axis. From this line cut we extract the transmon-resonator coupling energy $g$.
}
\label{fig:supplement-pinchoff}
\end{figure*}

We now investigate the performance of all electrostatic gates. 
Fig.~\ref{fig:supplement-pinchoff} shows the resonator frequency, measured by single-tone spectroscopy, while various combinations of gates are varied. In all cases, only at most one of the three junctions is open, thus not defining any loops. All junctions can be fully pinched off using any of the gates that control them while leaving the rest of the gates open, which confirms the proper functionality of all gates.

Panels a-c show the effect of varying the gates of ASQ1 either simultaneously (a) or separately (b, c). Note that the left and plunger gates of ASQ1 are connected to each other on-chip and thus are always set at the same voltage,~\Vlpa, while the right gate of ASQ1 is set at voltage~\Vra. 
The effect of the left, plunger and right gates of ASQ2, respectively set at voltages~\Vlb, \Vpb~and~\Vrb, is shown in panels e-h. Although the pinch-off voltages for ASQ2 are slightly lower, this junction displays a behavior similar to that of ASQ1.
Panel d shows the effect of~\Vc, the voltage of the coupling Josephson junction gate, which tunes $E_{\rm J,C}$. By varying~\Vc, the transmon frequency can be tuned to values above the bare resonator frequency, thus resulting in an avoided crossing between the resonator and transmon frequencies at around~\Vc~=~\SI{500}{mV}.
We find a transmon-resonator coupling strength $g/h\sim$~\SI{287}{MHz} as half of the distance between the two resonances observed at the avoided crossing.

Next, we measure the~\Vc-dependence of the transmon frequency $f_{\rm t}$ with both ASQs pinched off, from which we calibrate the \Vc~to~$E_{\rm J, C}$ map which is used for the data processing behind Fig.~4 in the main text. 

\begin{figure*}[h!]
\includegraphics{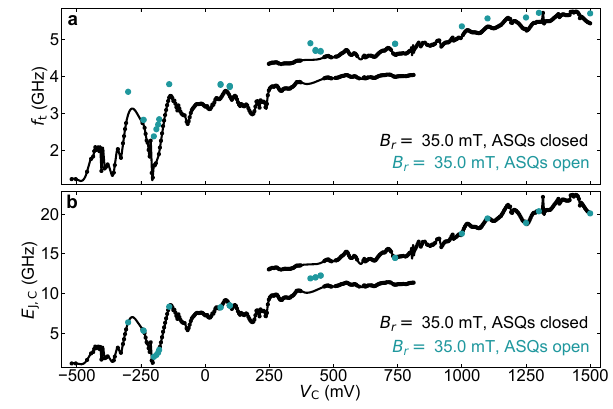}
\caption{{\bf Coupling junction characterization}
{\bf a} Transmon frequency, $f_{\rm t}$, versus the coupling junction gate voltage, \Vc, extracted by peak-finding in two-tone spectroscopy data. The 
black markers are taken at 
the magnetic field conditions at which we measured main text Fig.~4 but 
with both ASQ junctions fully closed (all quantum dot gates set to \SI{-1000}{mV}). The continuous line is a cubic interpolation to the measured data.  The teal markers indicate the \Vc, $f_{\rm t}$ points at which we measured longitudinal coupling in Fig.~4e. Note that, the teal markers deviate from the black markers, because there $f_{\rm t}$ is determined by the combination of $E_{\rm J, C}$, $E^I_{\rm J, 1}$, $E^I_{\rm J, 1}$, $E^\sigma_{\rm J, 1}$, $E^\sigma_{\rm J, 1}$, $\Phi_1$ and $\Phi_2$, and not solely by $E_{\rm J, C}$ as for the black markers.
{\bf b} Coupling junction Josephson energy  $E_{\rm J, C}$ versus \Vc~obtained directly from the corresponding data in {\bf a}, given the measured $E_{\rm c}$ value specified in Tab.~\ref{tab:circuit_parameters}. The mapping between \Vc~and  $E_{\rm J, C}$  indicated with a continuous black line is used to obtain the x-axis of Fig.~4e in the main text. 
}
\label{fig:supplement-transmon-EJ} 
\end{figure*}

The black markers in Fig.~\ref{fig:supplement-transmon-EJ}a show the transmon frequency as a function of \Vc, measured directly after the data shown in Fig.~4 of the main text and at the same magnetic field conditions ($B_r=$~\SI{35}{mT} applied in the chip plane and six degrees away from the $z$ direction). For comparison, we also show the transmon frequencies measured while taking the data in Fig.~4e (teal markers), with both ASQs open. In this case, the measured frequencies deviate from the black markers, since they instead result from a parallel combination of $E_{\rm J, C}$ and the Josephson energies of both qubits. The black markers in Fig.~\ref{fig:supplement-transmon-EJ}a are used to determine the \Vc-dependence of $E_{\rm J, C}$ shown in Fig.~\ref{fig:supplement-transmon-EJ}b, given the value of $E_{\rm c}$ independently determined from a measurement of the transmon anharmonicity (see Tab.~\ref{tab:circuit_parameters}). These data are used to determine the x-axis of Fig.~4e in the main text.

\subsection{ASQ gate setpoints} \label{Sss:setpoint}

In this section, we discuss the tune-up of each individual ASQs, which results in the chosen gate setpoints specified in Tab.~\ref{tab:setpoints}. 

\begin{figure}[h!]
    \center
    \includegraphics[scale=1.0]{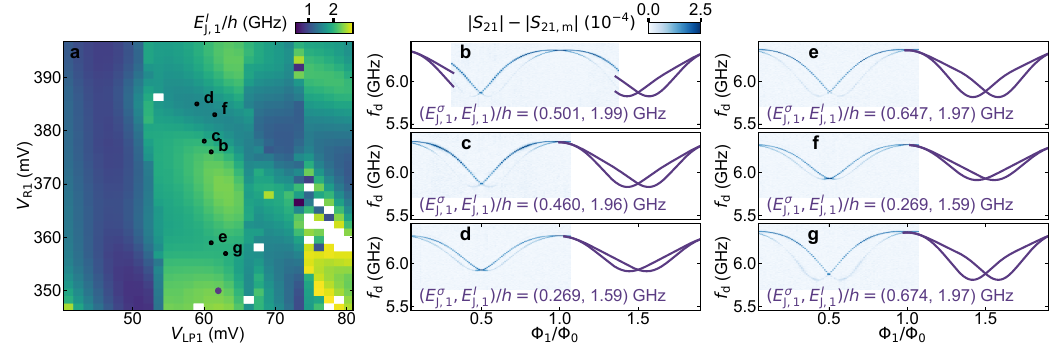}
    \caption{ {\bf ASQ1 gate dependence around its gate setpoint.} 
    {\bf a} Estimate of spin-independent Josephson energy of ASQ1, $E^I_{\rm J, 1}$, versus the two gate voltages of ASQ1, obtained from two-tone spectroscopy measurements of the transmon transitions at $B_{\rm r}=0$ and at two different flux values: $\Phi_1=0$ and $\Phi_1=\Phi_0/2$.
    The purple marker indicates the gate setpoint of ASQ1 in the main text.
    {\bf b} - {\bf g} Transmon spectroscopy at various gate configurations, indicated with markers in {\bf a}. The values of $E^I_{\rm J, 1}/h$  indicated in the labels are extracted by fitting the measured transmon frequencies at  $\Phi_1=0$ and $\Phi_1=\Phi_0/2$ to their theoretical values. The value of $E^\sigma_{\rm J, 1}/h$ is extracted from the distance between transmon frequencies at $\Phi_1=\Phi_0/4$. The continuous lines are the corresponding transmon frequencies obtained by numerical diagonalization of an adapted Eq.~\eqref{Eq:total-transmon-Hamiltonian} in which the spin-dependent potential of ASQ1 is replaced by a skewed sinusoidal shape $E^\sigma_{{\rm J}, 1} \sigma_1^z \sin{\left(\varphi_1+S\sin{\varphi_1}\right)}$, where $\varphi_1=\frac{2\pi}{\Phi_0}\Phi_1$. Panels {\bf b} - {\bf g}  share the color map. }
    \label{fig:ASQ1_2Dmap} 
\end{figure}

The tune-up of ASQ1 is presented in Fig.~\ref{fig:ASQ1_2Dmap}. We first set the junction containing ASQ2 to pinch-off by setting its three gates to \SI{-1000}{mV} and set the ASQ1 gates to a region where we detect a sizable spin-splitting energy in a low-resolution measurement.  From the transmon frequency at $\Phi_1=0,\Phi_0/2$ we estimate the spin-independent Josephson energy $E^I_{\rm J, 1}$ and map it out over a region in gate space using the two gate voltages of ASQ1 (Fig.~\ref{fig:ASQ1_2Dmap}a). Then, we proceed to investigating the value of the spin-dependent Josephson energy, $E^\sigma_{\rm J, 1}$. One way of doing so would be directly mapping out the spin-flip frequency $f_1$ in gate space. However, the visibility of the transition is significantly reduced at $B=0$ due to the thermal population of the ASQ as well as to the smaller matrix element from driving the spin transition. We instead perform $\Phi_1$-dependent transmon spectroscopy at a few selected gate points indicated with markers in Fig.~\ref{fig:ASQ1_2Dmap}a (Fig.~\ref{fig:ASQ1_2Dmap}b-g). 
For each gate setpoint we estimate the values of $E^\sigma_{\rm J, 1}/h$ by matching the distance between transmon frequencies at $\Phi_1=\Phi_0/4$ to its theoretically expected value extracted from numerical diagonalization of Eq.~\eqref{Eq:total-transmon-Hamiltonian} in the phase basis. Similarly, $E^I_{\rm J, 1}$ is estimated by fitting the measured transmon frequencies at $\Phi_1=0, \Phi_0/2$ to their theoretically expected values. The resulting quantities are indicated as labels on each panel. We choose the gate setpoint used for ASQ1 in the main text by maximizing $E^\sigma_{\rm J, 1}$ while keeping the value of $E^I_{\rm J, 1}$ low, since a high value negatively impacts the maximal coupling strength $J$. The chosen ASQ1 gate setpoint (see Tab.~\ref{tab:setpoints}) is indicated in Fig.~\ref{fig:ASQ1_2Dmap}a with a purple marker.

\begin{figure}[h!]
    \center
    \includegraphics[scale=1.0]{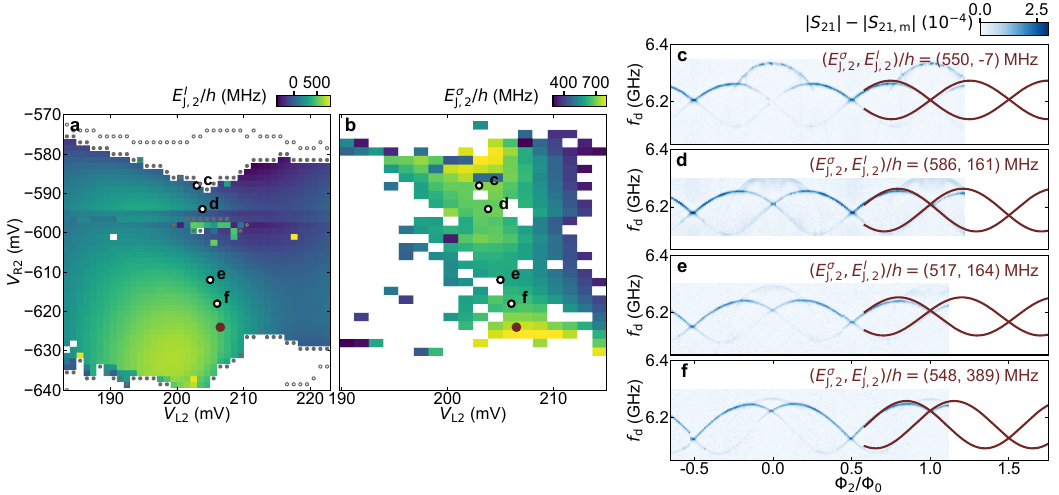}
    \caption{ {\bf ASQ2 gate dependence around its gate setpoint.} 
    {\bf a} Estimate of spin-independent Josephson energy of ASQ2, $E^I_{\rm J, 2}/h$, versus the two tunnel gates of ASQ2, for fixed \Vpb~=~0, obtained from two-tone spectroscopy measurements of the transmon transitions at $B_{\rm r}=0$ and at two different flux values: $\Phi_2=0$ and $\Phi_2=\Phi_0/2$.
    The open and filled grey markers indicate the boundaries of the singlet-doublet transition at $\Phi_2=0$ and $\Phi_2=\Phi_0/2$, respectively.
    {\bf b} Estimate of spin-dependent Josephson energy of ASQ2, $E^\sigma_{\rm J, 2}/h$, versus the two tunnel gates of ASQ2, for fixed \Vpb~=~0, obtained from two-tone spectroscopy measurements of the spin-flip transition, $f_2$, at $B_{\rm r}=0$ and $\Phi_2=\Phi_0/4$.
    The maroon marker in {\bf a} and {\bf b} indicates the gate setpoint of ASQ2 in the main text.
    {\bf c} - {\bf f} Transmon spectroscopy at various gate configurations indicated with markers in {\bf a} and {\bf b}. The values of $E^I_{\rm J, 2}/h$  indicated in the labels are extracted by fitting the measured transmon frequencies at  $\Phi_2=0$ and $\Phi_2=\Phi_0/2$ to their theoretical values. The value of $E^\sigma_{\rm J, 2}/h$ is extracted from the distance between transmon frequencies at $\Phi_2=\Phi_0/4$. The continuous lines are the corresponding transmon frequencies obtained by numerical diagonalization of Eq.~\eqref{Eq:total-transmon-Hamiltonian}. Panels {\bf c} - {\bf f}  share the color map.
    }
    \label{fig:ASQ2_2Dmap} 
\end{figure}

\begin{figure}[h!]
    \center
    \includegraphics[scale=1.0]{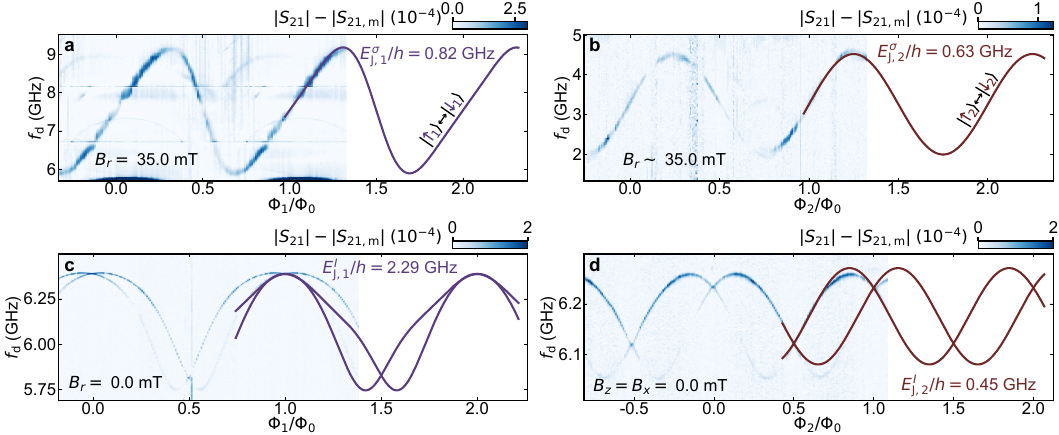}
    \caption{ {\bf Parameter estimation for both ASQs.} 
    {\bf a} Spin-flip spectroscopy of ASQ1 versus $\Phi_1$, at $B_r$~=~\SI{35}{mT}.  The line shows a fit to a skewed sinusoidal dependence (Eq.~\eqref{eq:sin_sk}) from which we extract the value of $E^\sigma_{\rm J, 1}/h$~=~\SI{0.82}{GHz}.
    {\bf b} Spin-flip spectroscopy of ASQ2 versus $\Phi_2$, at $B_r \sim$~\SI{35}{mT}.  The line shows a fit to a sinusoidal dependence (Eq.~\eqref{eq:sin}) from which we extract the value of $E^\sigma_{\rm J, 2}/h$~=~\SI{0.63}{GHz}.
    {\bf c} Transmon spectroscopy versus $\Phi_1$, at $B_r$~=~\SI{0}{mT} with ASQ1 open to its setpoint (see Tab.~\ref{tab:setpoints}) and ASQ2 closed. The two transmon frequencies correspond to the two possible states of ASQ1. 
    {\bf d} Transmon spectroscopy versus $\Phi_2$, at $B_r$~=~\SI{0}{mT} with ASQ2 open to its setpoint (see Tab.~\ref{tab:setpoints}) and ASQ1 closed. The two transmon frequencies correspond to the two possible states of ASQ2. For {\bf a} and {\bf b}, \Vc~=~\SI{1500}{mV}, while for {\bf c} and {\bf d}, \Vc~=~\SI{1995}{mV}.
    The continuous lines in {\bf c} and {\bf d} show the transmon transition spectrum given the spin-dependent part of the ASQ potentials found in {\bf a}, {\bf b} and the measured value of $E_{\rm c}$ (see Tab.~\ref{tab:circuit_parameters}). In both cases, the frequencies are obtained by numerical diagonalization of the Hamiltonian in Eq.~\eqref{Eq:total-transmon-Hamiltonian} and are best fits of the measured data at $\Phi_i$ being integer multiples of $\Phi_0/2$. From these transmon spectra, we extract the values of the spin-independent Josephson energies $E^I_{\rm J, 1}/h$~=~\SI{2.29}{GHz} and $E^I_{\rm J, 2}/h$~=~\SI{0.45}{GHz}. 
    }
    \label{fig:parameter-extraction} 
\end{figure}

Next, we pinch off the junction containing ASQ1 to tune-up the gate configuration of ASQ2. We perform an investigation analogous to the one detailed above, as shown in Fig.~\ref{fig:ASQ2_2Dmap}.  Fig.~\ref{fig:ASQ2_2Dmap}a is measured in the same way as  Fig.~\ref{fig:ASQ1_2Dmap}a and displays the evolution of $E^I_{\rm J, 2}$ with the tunnel gates, \Vlb~and \Vrb, while \Vpb~is kept at \SI{0}{mV}. Fig.~\ref{fig:ASQ2_2Dmap}b shows the tunnel gate dependence of $E^\sigma_{\rm J, 2}$, determined from direct spin-flip spectroscopy of ASQ2 at $B=0$: $E^\sigma_{\rm J, 2}/h = f_2(B=0, \Phi_2=\Phi_0/4)/2$. Similarly to the strategy for ASQ1, we choose a gate setpoint for ASQ2 by maximizing $E^\sigma_{\rm J, 2}$ while keeping $E^I_{\rm J, 2}$ as low as possible. However, for some gate points in this region of gate space, a singlet state is also slightly visible in transmon spectroscopy (as can be seen in Fig.~\ref{fig:ASQ2_2Dmap}c, around $\Phi_2=0$). The presence of the singlet state indicates that the singlet phase of the system is only separated by an energy gap comparable to the thermal energy of the system. Consequently, while choosing the ASQ2 setpoint we also minimize the visibility of the singlet state. The chosen setpoint (see Tab.~\ref{tab:setpoints} is indicated in Fig.~\ref{fig:ASQ2_2Dmap}a and b with a maroon marker).

Finally, we perform in-field spin-flip spectroscopy of both ASQs, as well as transmon spectroscopy at zero field, to more accurately determine their Josephson energies detailed in Tab.~\ref{tab:setpoints}. 

\begin{table*}[h!]
\begin{ruledtabular}
\begin{tabular}{cccccc}
\textrm{}& $V_{{\rm L} i}$ (mV)  & $V_{{\rm P} i}$ (mV)  & $V_{{\rm R} i}$ (mV) & $E^I_{{\rm J}, i}/h$ (GHz) & $E^\sigma_{{\rm J}, i}/h$ (GHz) \\
\colrule
ASQ1 & 62.0 & 62.0 & 350.0 & 2.29 & 0.82  \\
ASQ2 & 206.50 & 0.0 & -624.0 & 0.45 & 0.63  \\
\end{tabular}
\end{ruledtabular}
\caption{
ASQ1 and ASQ2 gate voltage set points and extracted model parameters from the measurements in Fig.~\ref{fig:parameter-extraction}.
}
\label{tab:setpoints}
\end{table*}

The spin-flip spectroscopy shown in Fig.~\ref{fig:parameter-extraction}a and b is performed under the same magnetic field conditions as the ones where we measured coupling in the main text. We extract $E^\sigma_{{\rm J}, 1}/h=$~\SI{0.82}{GHz} and $E^\sigma_{{\rm J}, 2}/h=$~\SI{0.63}{GHz} from fits of a skewed and non-skewed sine, respectively, to the measured data (see Sec.~\ref{sss:fits_spin_flip}). The $E^\sigma_{{\rm J}, i}/h=$ values are determined as one fourth of the flux dispersion of the fit result. $E^I_{{\rm J}, 1}/h=$~\SI{2.29}{GHz} and $E^I_{{\rm J}, 2}/h=$~\SI{0.45}{GHz} are determined similarly to how it was done for Figs.~\ref{fig:ASQ1_2Dmap} and \ref{fig:ASQ2_2Dmap}, by fitting the $\Phi_i$-dependent data to the expected transmon frequencies obtained by numerical diagonalization of the Hamiltonian in Eq.~\eqref{Eq:total-transmon-Hamiltonian} at $\Phi_i$ being integer multiples of $\Phi_0/2$. In both cases, we fix the spin-dependent part of the transmon potential to that extracted from the fits in Fig.~\ref{fig:parameter-extraction}a and b.

\subsection{Andreev spin qubit readout} \label{Sss:readout}

\begin{figure}[h!]
    \center
    \includegraphics[scale=1.0]{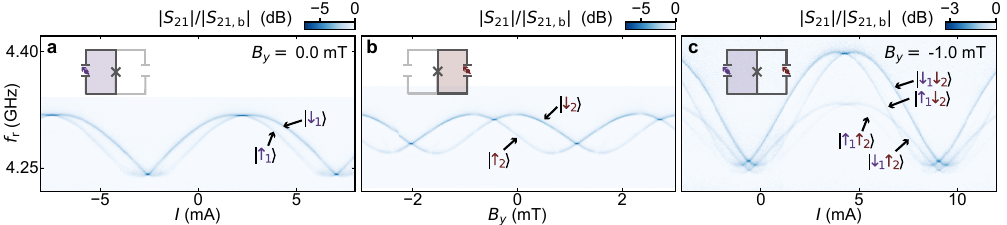}
    \caption{ {\bf Readout.} {Amplitude of the transmission through the readout circuit, $|S_{21}|$, with background, $|S_{21, \rm b}|$, divided out. }
    {\bf a} With ASQ1 open at its setpoint (see Tab.~\ref{tab:setpoints}) and ASQ2 closed (\Vlb~=~\Vpb~=~\Vrb~=~\SI{-1000}{mV}). 
    {\bf b} With ASQ1 closed (\Vlpa~=~\Vra~=~\SI{-1000}{mV}) and ASQ2 open  at its setpoint (see Tab.~\ref{tab:setpoints}). 
    {\bf c} With both ASQ1 and ASQ2 open, as also displayed in Fig.~1d of the main text.
    {\bf a, c} are plotted vs. the current through the flux line $I$, which controls $\Phi_{1}$. {\bf b}  is plotted vs. $B_y$, which controls  $\Phi_{2}$. 
    For all panels, $B_x = B_z = 0$ and \Vc~=~\SI{180}{mV}.
    }
    \label{fig:readout} 
\end{figure}

In the main text (Fig.~1d) we discussed how, when both loops are open, we observe four possible resonator frequencies, depending on the four possible spin states of the ASQ1-ASQ2 system, $\left\{\ket{\uparrow_1\uparrow_2}, \ket{\uparrow_1\downarrow_2},\ket{\downarrow_1,\uparrow_2},\ket{\downarrow_1\downarrow_2}\right\}$. This allows us to perform two-tone spectroscopy of either one of the two qubit transitions, $f_1$ and $f_2$, which are present when both ASQ junctions are open. Here, we show the analogous situation when only {\it one} out of the two ASQs is open, while the junction containing the other one is fully pinched off (Fig.~\ref{fig:readout}).

Fig.~\ref{fig:readout}a shows the $\Phi_1$-dependence of resonator spectroscopy, at zero magnetic field and when only ASQ1 is open. In this case, we observe two branches of the resonator frequency, corresponding to the two possible states of ASQ1: $\ket{\uparrow_1}$ or $\ket{\downarrow_1}$). The different visibility of each of the branches is a consequence of the different thermal populations of the two states at $B_r=0$. This is expected, since the spin-splitting of ASQ1 varies with flux reaching up to $2E^\sigma_{{\rm J}, 1}/h=$~\SI{1.64}{GHz}, comparable, when transformed into an effective temperature, to typical electron temperatures found in other experiments \cite{Jin2015, PitaVidal2023, Uilhoorn2021}. 
Fig.~\ref{fig:readout}b shows the analogous situation but now for ASQ2. In this case, the resonator also displays two separate frequencies. After fixing $B_y$ so that $\Phi_2\sim - \Phi_0/4$ and so that the separation between the resonator frequencies corresponding to $\ket{\uparrow_2}$ and $\ket{\downarrow_2}$ is sizable, we open ASQ1 to its setpoint. In such situation, when performing resonator spectroscopy versus $\Phi_1$, we observe four different transitions, labeled with their corresponding states in Fig.~\ref{fig:readout}c. Note that, in this case, the difference in visibility becomes more perceptible due to the larger energy separation between the different states $\left\{\ket{\uparrow_1\uparrow_2}, \ket{\uparrow_1\downarrow_2},\ket{\downarrow_1,\uparrow_2},\ket{\downarrow_1\downarrow_2}\right\}$.

\subsection{Magnetic field angle dependence and determination of the spin-orbit direction} \label{Sss:angle}

\begin{figure}[h!]
\includegraphics{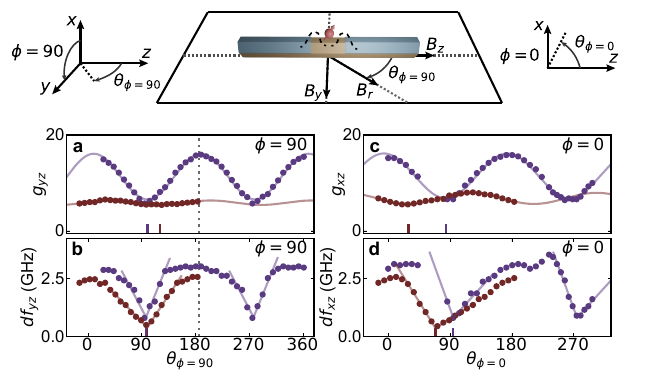}
\caption{{\bf Magnetic field angle dependence.}
{\bf a} $g$-factor for both ASQs for varying magnetic field direction in the chip plane, $y-z$, plotted as a function of the angle between the applied field and the nanowire axis, $\theta_{\phi=90}$. The $g$-factor is calculated as the average between its maximum and minimum values versus flux. Data points corresponding to ASQ1 and ASQ2 are colored purple and maroon, respectively. The continuous lines are cosinusoidal fits to the data. 
{\bf b} Flux-dispersion of the spin-flip, $df_{yz}$, for both ASQs versus $\theta_{\phi=90}$. $df_{yz}$ is calculated as the difference between the maximum and minimum of the spin-flip frequency $f_i$ versus flux. The continuous lines are fits to the data around their minima and the colored vertical lines on the x-axis indicate the positions of the minima extracted from the fits, which are interpreted as the directions perpendicular to the zero-field spin-polarization direction for each qubit. Note that these lines do not coincide with the minima of the $g$-factors found in {\bf a}.
The vertical dotted lines in {\bf a}, {\bf b} indicate the field angle along which all measurements in the main text, except for Fig.~2c,d, were taken.
{\bf c}, {\bf d} Same as {\bf a}, {\bf b} but in the $x-z$ plane, the plane perpendicular to the chip which contains the nanowires axis.
}
\label{fig:supplement-angle-dep} 
\end{figure}

In this section, we specify the measurements performed to determine the zero-field spin-polarization direction for each Andreev spin qubit. For each qubit, we perform spin-flip spectroscopy measurements, like those shown in Fig.~\ref{fig:parameter-extraction}a and b, for different magnetic field directions. As reported previously in Ref.~\cite{Bargerbos2022b}, we observe that both the flux dispersion of the spin-flip transition, $df$, as well as the $g$-factor, depend strongly on the direction of the applied magnetic field. To determine these quantities, the maxima, $f_i^{\rm max}$, and minima, $f_i^{\rm min}$, of the spin-flip frequencies are first extracted by hand  from two-tone spectroscopy measurements of the spin-flip transition,  analogous to those in Fig.2a-c of the main text. The $g$-factors are calculated from the average of these maximum and minimum frequencies, as $g = (f_i^{\rm max}+f_i^{\rm min})/(2\mu_{\rm B} B_r)$, where $\mu_{\rm B}$ is the Bohr magnetron and $B_r$ is the magnitude of the applied magnetic field. The frequency dispersion is determined as $df=(f_i^{\rm max}-f_i^{\rm min})/2$. The dependence of $g$ and $df$ on the magnetic field direction is shown in  Fig.~\ref{fig:supplement-angle-dep} with purple and maroon markers for ASQ1 and ASQ2, respectively. 

First, we investigate the dependence on the angle within the chip plane and away from the nanowires axis, $\theta_{\phi=90}$.  $\theta_{\phi=90}=0$ indicates that the field is applied approximately along the nanowires axis, while 
$\theta_{\phi=90}=90$ degrees indicates that the field is applied in-plane but approximately perpendicular to the nanowire axis. We find that the $g$-factor of ASQ1 depends strongly on $\theta_{\phi=90}$, while that of ASQ2 stays almost constant, fluctuating only between 5.5 and 6.5 (Fig.~\ref{fig:supplement-angle-dep}a). Within this plane, the $g$-factor of ASQ1 is found to be maximal when the magnetic field $B_r$ is applied approximately along the nanowires axis, while for ASQ2 it is maximized for $\theta_{\phi=90}\sim 31$ degrees away from the nanowire axis. Performing the same experiment while varying the field direction in the $x$-$z$ plane, the plane perpendicular to the chip and containing the nanowires axis, we observe a similar dependence (Fig.~\ref{fig:supplement-angle-dep}c). This time, the ASQ1 $g$-factor is again maximized along the nanowires axis, while that of ASQ2 becomes maximal when $B_r$ is applied $\theta_{\phi=0}\sim 60$~degrees away from the nanowires axis. This variability of the $g$-factor dependence for different configurations is consistent with previous observations of quantum dots implemented in InAs nanowires and is thought to be due to mesoscopic fluctuations of the electrostatic environment at the quantum dot \cite{Han2023, Bargerbos2022b}.

To learn about the zero-field spin-polarization direction of each qubit, we now focus on the field-direction dependence of the flux dispersion of the spin-flip transition. We denote by $df$ the difference in frequency between the maximum and minimum of the spin-flip frequency versus flux. When the field is applied along the zero-field spin-polarization direction, we expect that $df=4 E^\sigma_{{\rm J}, i}$ (see Sec.~\ref{sec:theory}). However, when a component of the applied magnetic field is perpendicular to the zero-field spin-polarization direction, $df$ is reduced due to the hybridization of the two spin states \cite{Bargerbos2022b}. Fig.~\ref{fig:supplement-angle-dep}b shows the $\theta_{\phi=90}$ dependence of $df$ for both ASQs. We find that, in both cases, $df$ becomes minimal at a direction approximately perpendicular to the nanowire axis.
We also perform a similar experiment in the $x$-$z$ plane (see Fig.~\ref{fig:supplement-angle-dep}d). 
The extracted angles
constitute two of the directions perpendicular to the spin-polarization axis. Their cross-product thus returns the zero-field spin polarization directions for each qubit, which are indicated in Tab.~\ref{tab:directions} in spherical coordinates, where $\theta_{||}$ denotes the polar angle away from the nanowire axis and  $\phi_{||}$ denotes the azimuthal angle measured away from the $x$-axis (see Fig.~\ref{fig:supplement-angle-dep}). The angle between this direction and the direction of the applied magnetic field in Figs.~3 and 4 of the main text ($\theta=185, \phi=90$) is indicated in the last column of Tab.~\ref{tab:directions}.

The angle used for all measurements in the main text, except for Fig.~2c,d,  is indicated with vertical dotted lines in Fig.~\ref{fig:supplement-angle-dep}a and b. We chose this angle following various considerations. First, we wanted to minimize the field component perpendicular to the spin directions of each of the ASQs. The reason for this is that we expect transverse qubit-qubit coupling terms to emerge under the presence of a large perpendicular Zeeman energy compared to $E_{\rm J, i}^\sigma$, at the cost of the longitudinal component. Second, we wanted to maximize the difference in $g$-factors to avoid crossings between the qubit frequencies versus flux. This enables the possibility of spectroscopically measuring the coupling strength at every flux point. Finally, we chose the total field magnitude $B_r=$~\SI{35}{mT}  to set the ASQ frequencies at setpoints that did not cross neither the resonator nor any transmon transition frequency for any value of flux.

\begin{table}[h!]
\begin{ruledtabular}
\begin{tabular}{cccc}
\textrm{}&  $\theta_{||}$  & $\phi_{||}$ & $\alpha$ \\
\colrule
ASQ1 & 8.54 & 54.76 & 5.1 \\
ASQ2 & 22.73 & 157.15 & 21.5 \\
\end{tabular}
\end{ruledtabular}
\caption{ {\bf Zero-field spin-polarization directions for ASQ1 and ASQ2 in degrees.} 
 The zero-field spin-polarization direction ($\theta_{||}, \phi_{||}$) is calculated as the vector product of the two perpendicular directions 
 indicated with colored lines in the x-axis of Fig.~\ref{fig:supplement-angle-dep}b,d. 
 $\alpha$ denotes the angle between the field applied in the main text measurements and the spin-polarization direction for each qubit.
}\label{tab:directions}
\end{table}

\FloatBarrier
\newpage
\newpage
\section{\label{coherence-data} Supplementary coherence data}
We now describe the functions used for extracting the coherence times quoted in the main text. To determine  $T_1$  we fit an exponential decay 
\begin{equation}\label{eq:coh-T_1_decay}
    y(t) = a\cdot \mathrm{exp}[{t/T_1}] + c
\end{equation}
where $a$, $T_1$ and $c$ are free fit parameters.
For Ramsey and Hahn echo (see~\cref{fig:supplement-T2echo}) experiments we fit a sinusoide with a exponential decay envelope and sloped background 
\begin{equation}\label{eq:coh-T_2_decay}
 y(t) = a \cdot \cos\left(\frac{2\pi}{ p} t - \phi\right)\cdot\mathrm{exp}\left[(-t/T_2)^{d+1}\right] + c + et 
 \end{equation}
 where $a$, $T_2$, $\phi$, $c$, $e$ and $p$ are free fit parameters and $d$ was fixed to 1, resulting in a Gaussian decay envelope. We found that $d=1$ gave the least $\chi$-squared in the fit compared to $d=0,2$. The tilted background was included to compensate for a slightly non-linear Rabi response. 

\subsection{Hahn echo decay time measurements of ASQ1 and ASQ2}\label{sec:hahn_decay}
\begin{figure*}[h!]
\includegraphics{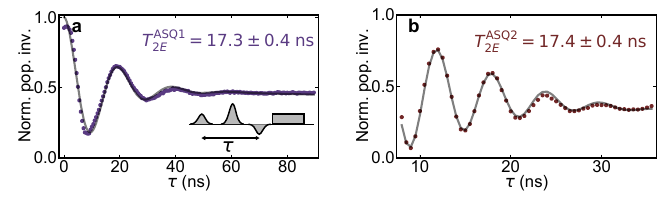}
\caption{{\bf Hahn echo experiment on ASQ1 and ASQ2.}
{\bf a, b} Measurement of $T_2$-echo for ASQ1 and ASQ2, respectively, with artificial detuning at the setpoints indicated in Fig.~2 of the main text. The pulse sequence is shown in the inset of {\bf a}. This is the same sequence as used in the Ramsey experiment, but with a $\pi$-pulse added between the two $\pi/2$ pulses. In {\bf a}, an artificial detuning corresponding to a period of \SI{20}{\nano\second} was set and in {\bf b} it was set to \SI{6}{\nano\second}. This data was taken using Gaussian pulses with FWHM of \SI{4}{\nano\second} and averaged over $3\cdot 10^5$ shots for each data point. The y-axis is normalized using the fit (for details, see the accompanying text in~\cref{sec:hahn_decay}). 
}
\label{fig:supplement-T2echo} 
\end{figure*}

To verify the slow nature of the noise causing dephasing, we performed Hahn-echo experiments with artificial detuning, shown in~\cref{fig:supplement-T2echo}. The resulting data was fit using~\cref{eq:coh-T_2_decay}. Note that for these measurement we found that the data was not always within the range of the identity and $\pi$-pulse calibration points. We suspect this is due to the additional echo pulse inducing leakage to other states outside the spin-subspace. We therefore normalized the data setting 0 and 1 to the minimum and maximum of the fit envelope at $\tau=0$ instead of using the calibration points as was done in the main text.  

\subsection{Coherence properties of the transmon}
\begin{figure*}[h]
\includegraphics{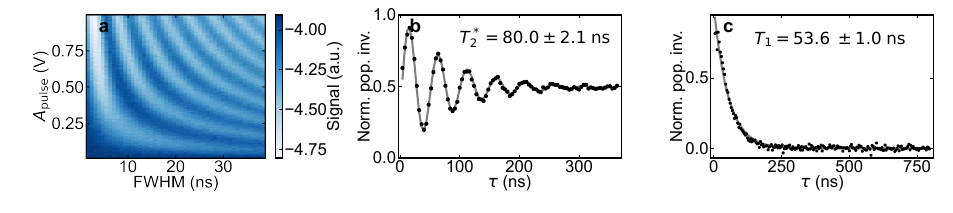}
\caption{{\bf Transmon coherence properties.}
{\bf a} Rabi oscillations versus the full-width at half maximum (FWHM) and amplitude of the Gaussian pulse. 
{\bf b} Standard $T_2^*$ measurement using a Ramsey sequence with Gaussian pulses of FWHM~$=\SI{5.5}{n s}$. The fit to~\cref{eq:coh-T_2_decay} (grey line) was performed here with $d=0$. 
{\bf c} $T_1$ measurement. 
For panels {\bf b} and {\bf c} the data was normalized using the fitted scaling constants of~\cref{eq:coh-T_1_decay,eq:coh-T_2_decay}.
}
\label{fig:supplement-tmon-coherence} 
\end{figure*}

Although the transmon was used in this work to facilitate spin readout, we now demonstrate its coherence properties. \cref{fig:supplement-tmon-coherence} shows measurements of the transmon's Rabi oscillations, Ramsey coherence time and energy decay time. 

The transmon $T_1$ was found to be lower than that of previous implementations of a transmon using gate-tunable nanowires \cite{Larsen2015, Casparis2016, Luthi2018, Kringhoj2021, Bargerbos2023, PitaVidal2023}, which we suspect may be due to it being too strongly coupled to the flux-bias line or drive lines. This in turn limits the ASQs $T_1$ due to Purcell decay via the transmon.  Given a transmon $T_1$ of \SI{53.6}{ns}, we can estimate the limit set by Purcell decay for each ASQ. At their setpoints in Fig.~2 in the main text, the detunings from the transmon were \SI{1.7}{GHz} and \SI{2.2}{GHz} for ASQ1 and ASQ2, respectively. From measurements of the avoided crossing between the ASQ spin-flip and transmon transitions under similar conditions, we extract that the coupling strengths between transmon and ASQ are in the range \SIrange{50}{100}{MHz} for both qubits. These quantities allow us to estimate the Purcell limit of $T_1$ for both qubits to be 14-\SI{56}{\micro s} for ASQ2 and  23-\SI{92}{\micro s} for ASQ1. The higher harmonics of the transmon can reduce these lifetimes even more, especially for ASQ1 which resides close to the first higher harmonic. However, we did not measure the lifetime of that transition.

\FloatBarrier

\subsection{Scaling of extracted T2* with pulse length}
\begin{figure*}
\includegraphics{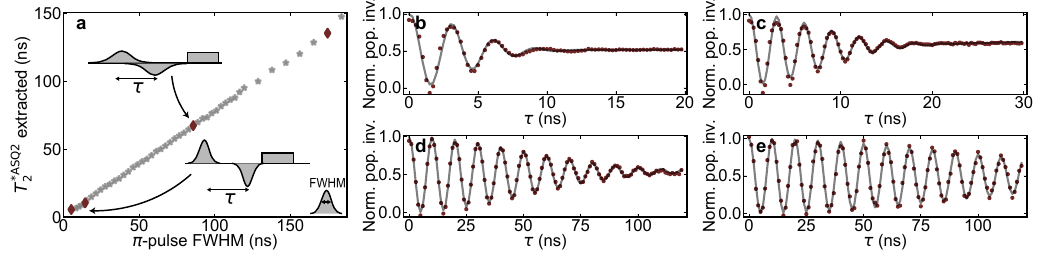}
\caption{{\bf Effect of overlapping pulses on the extracted $T_2^{*{\rm ASQ2}}$ for ASQ2. } {\bf a} Extracted $T_2^{*{\rm ASQ2}}$ by performing Ramsey experiments on ASQ2 as a function of the  $\pi/2$-pulse FWHM. For each data point we reduced the amplitude of the $\pi/2$-pulse by the same factor as we increased the pulse length to keep an approximate $\pi/2$-pulse, using the calibration for the shortest pulse length. {\bf b}-{\bf d} Examples of the data with $T_2^{*{\rm ASQ2}}$ fits indicated with maroon markers in {\bf a}. The artificial detuning period was varied with pulse length to make sure there were enough points in each period and enough oscillations in the length of the traces.}
\label{fig:supplement-pulse-length-dep} 
\end{figure*}

Due to the short dephasing time of ASQs with respect to the pulse length, the pulse length influences the observed life time of the ASQs when the pulses are (partly) overlapping (see ~\cref{fig:supplement-pulse-length-dep}). This is the case because, during the part of the decay time $\tau$, the ASQ is being driven.  Therefore, care should be taken when pulses of length comparable to $T_2^*$ are used. In the main text we report values obtained using short \SI{4}{\nano\second} FWHM pulses. 

\subsection{Single-shot readout contrast of individual ASQs}
\begin{figure*}
\includegraphics{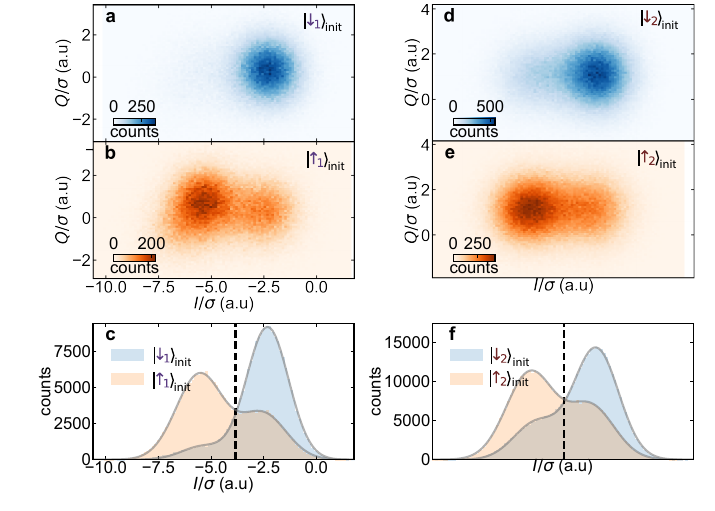}
\caption{{\bf Individual qubit readout shots}.  {\bf a}-{\bf c} Histograms of single shot readout of ASQ1 near the maximum of the transition ($I=\SI{-2}{\milli\ampere}$) at $B_z=\SI{35}{\milli\tesla}$ (same setpoint as for the coherence measurements in the main text). {\bf a}, {\bf b} Histogram of single shot  measurements in the $I-Q$ plane after initializing in $\ket{\downarrow_1}$ and $\ket{\uparrow_1}$, respectively. {\bf c} Projection of the data in {\bf a} and {\bf b} along the $I$-axis fitted with a double Gaussian function (grey line) for each initialization. A \SI{1000}{\nano\second} readout pulse and a $\sim$\SI{2}{\nano\second} FWHM $\pi$-pulse were used.
The black dashed line in {\bf c} indicates the optimal threshold to distinguish spin-up from spin-down states and is used to calculate the fidelity $F$.  {\bf d}-{\bf f} Histograms of single-shot readout of ASQ2 at $B_z=\SI{80}{\milli\tesla}$ for a \SI{1500}{\nano\second}-long readout pulse and a $\sim\SI{4.7}{\nano\second}$ FWHM $\pi$-pulse, at $B_y=\SI{5.96}{\milli\tesla}$.  }
\label{fig:supplement-Readout-fidelity} 
\end{figure*}

In~\cref{fig:supplement-Readout-fidelity} we show examples of single-shot readout outcomes. These are measured at the setpoints used in the main text and at magnetic field settings of the main text for ASQ1 and for ASQ2 we go to a higher magnetic field in order to reduce thermal population of the excited spin state. We obtain an average signal-to-noise ratio for distinguishing spin-up and spin-down, based on double Gaussian fits to the up and down initializations $\textrm{SNR}=|\mu_\uparrow-\mu_\downarrow|/(2\sigma)$ where $\mu_i, \sigma$ are the mean and width of the fitted Gaussian corresponding to state $i$, of 1.5 and 1.3 in a integration time of $\SI{1}{\micro\second}$, $\SI{1.5}{\micro\second}$ for ASQ1 and ASQ2 respectively. Note that we use the fit parameters of the initialization without a $\pi$-pulse here as the pulse can cause excitations of other states, which we suspect to be the transmon excited states, visible as an extra tail in the Gaussian corresponding to the excited spin state in~\cref{fig:supplement-Readout-fidelity}b, c. Additionally, these values are strongly flux and magnetic field dependent and thus could be optimized further in future works, as we did not perform an exhaustive study here.  

The SNR is a  measurement of the pure readout contrast, rather than the more standard readout fidelity $F=1-P(\uparrow|\downarrow)/2 - P(\downarrow|\uparrow)/2$, where $P(a|b)$ denotes the probability of obtaining state $a$ when preparing state $b$. This is because $F$ includes the effects of thermal population and imperfect $\pi$-pulse, due to dephasing during the $\pi$-pulse and imperfect calibrations. At the main text gate setpoint, and the magnetic field settings mentioned above using the indicated threshold (black dashed line in~\cref{fig:supplement-Readout-fidelity}) we obtain $F=0.75$ and $F=0.67$ for ASQ1 and ASQ2 respectively.

\FloatBarrier

\newpage
\newpage
\section{Supporting data for the longitudinal coupling measurements}\label{Sss:consistency-checks}

\begin{figure}[h!]
    \center
    \includegraphics[scale=1.0]{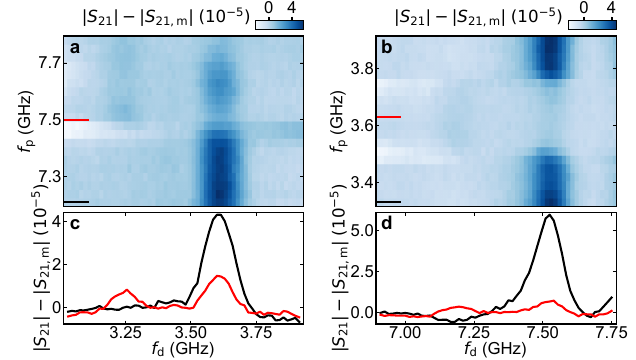}
    \caption{ {\bf Third tone frequency dependence of the longitudinal coupling signal.} 
    {\bf a} Spectroscopy of ASQ2 as a function of the drive frequency $f_{\rm d}$ while continuously applying a pump tone at varying frequency $f_{\rm p}$. The red line indicates the pump tone frequency used in Fig.~3b,d of the main text, $f_{\rm p} = f_1-J$.
    {\bf b}  Same as in {\bf a} but with the roles of ASQ1 and ASQ2 exchanged. In this case, the pump tone has a frequency close to that of ASQ2, while performing spectroscopy of ASQ1. 
    {\bf c}  and {\bf d} show line-cuts of {\bf a} and {\bf b}, respectively, at two fixed pump frequencies indicated in {\bf a} and {\bf b} with color matching lines.
    }
    \label{fig:consistency_check_fp} 
\end{figure}

Fig.~\ref{fig:consistency_check_fp} shows data taken in the same way as in Fig.~3 of the main text and under the same field, gate and flux conditions, but for varying frequency of the pump tone $f_{\rm p}$. We find that, when $f_{\rm p}$ matches the transition frequency of one of the qubits, and thus continuously drives it to its excited state, the frequency of the other qubit splits in two, as discussed in the main text (red lines in Fig.~\ref{fig:consistency_check_fp}c and d). When the pump tone frequency instead does not match the transition frequency of the first qubit, the frequency of the second qubit does not split, as expected (black lines in Fig.~\ref{fig:consistency_check_fp}c and d). This confirms that the frequency splitting observed in the main text is indeed the result of both states of the other ASQ being populated.

\begin{figure}[h!]
    \center
    \includegraphics[scale=1.0]{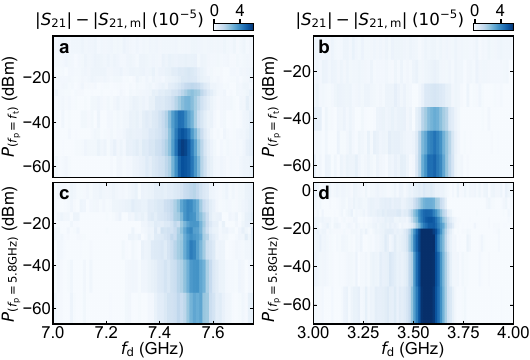}
    \caption{ {\bf Qubit spectroscopy while driving at different frequencies with the third pump tone.} 
    {\bf a} Two-tone spectroscopy of ASQ1 while driving with the third pump tone at the transmon frequency $f_{\rm p} = f_{\rm t}$ with varying power.
    {\bf c} Same as {\bf a} but for $f_{\rm p} = 5.8$~GHz, not in resonance with any visible transition.
    {\bf b} and {\bf d} Same as {\bf a} and {\bf c} but for ASQ2.
    }
    \label{fig:consistency_check_frequencies} 
\end{figure}

In Fig.~\ref{fig:consistency_check_frequencies} we perform a similar experiment for two fixed pump frequencies away from the spin-flip transitions and as a function of the  pump tone power. For Fig.~\ref{fig:consistency_check_frequencies}a and b, the pump tone drives the transmon transition and for  Fig.~\ref{fig:consistency_check_frequencies}c and d its frequency is set to a value $f_{\rm p}=\SI{5.8}{\giga\hertz}$ where no transition is driven. In neither of the two cases do we observe a splitting of any of the two ASQ transitions at any power, as expected (the disappearance of the signal at high drive powers was generally seen throughout the work independent of drive frequency and corresponds to disappearance of the readout resonator resonance). Note that the instability that can be observed in the measured transition frequencies was also observed versus time and is thus unrelated to the presence of the pump tone.

\section{Longitudinal coupling at different gate sepoint} \label{s:new-dataset}

In this section, we present longitudinal coupling measurements similar to those in the main text, but now taken at a different gate configuration for both Andreev spin qubits. The new gate setpoints, at which the two qubits are set for all results discussed in this section, are indicated in Tab.~\ref{tab:setpoints_new}. 

\begin{table*}[h!]
\begin{ruledtabular}
\begin{tabular}{cccccc}
\textrm{}& $V_{{\rm L} i}$ (mV)  & $V_{{\rm P} i}$ (mV)  & $V_{{\rm R} i}$ (mV) & $E^I_{{\rm J}, i}/h$ (GHz) & $E^\sigma_{{\rm J}, i}/h$ (GHz) \\
\colrule
ASQ1 & 61.0 & 61.0 & 376.0 & 1.79 & 0.66  \\
ASQ2 & 53.0 & 0.0 & -700.0 & 0.53 & 1.3  \\
\end{tabular}
\end{ruledtabular}
\caption{
ASQ1 and ASQ2 gate voltage set points and extracted model parameters from the measurements in Fig.~\ref{fig:parameter-extraction_new} .
}
\label{tab:setpoints_new}
\end{table*}

\begin{figure}[h!]
    \center
    \includegraphics[scale=1.0]{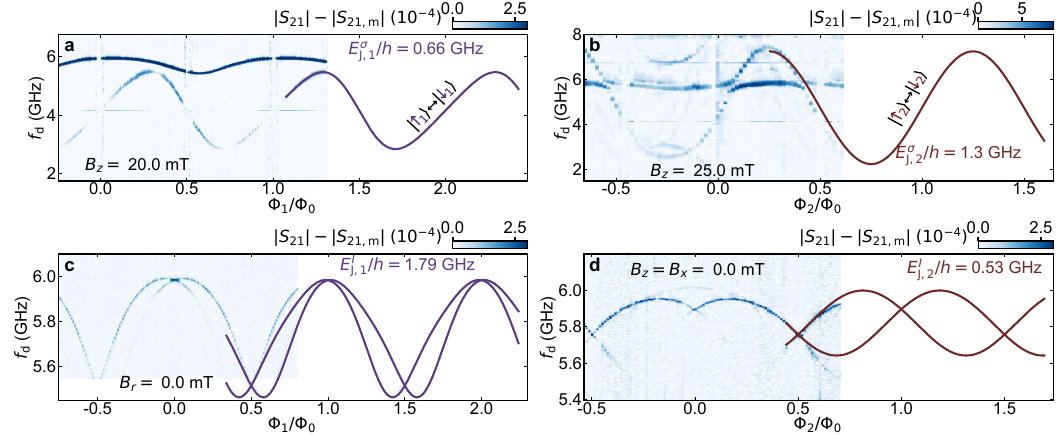}
    \caption{ {\bf Parameter estimation for both ASQs at the gate setpoints specified in Tab.~\ref{tab:setpoints_new}.} 
    {\bf a} Spin-flip spectroscopy of ASQ1 versus $\Phi_1$, at $B_z$~=~\SI{20}{mT}.  The line shows a fit to a skewed sinusoidal dependence (Eq.~\ref{eq:sin_sk}) from which we extract the value of $E^\sigma_{\rm J, 1}/h$~=~\SI{0.66}{GHz}.
    {\bf b} Spin-flip spectroscopy of ASQ2 versus $\Phi_2$, at $B_z \sim$~\SI{25}{mT}.  The line shows a fit to a sinusoidal dependence (Eq.~\ref{eq:sin}) from which we extract the value of $E^\sigma_{\rm J, 2}/h$~=~\SI{1.3}{GHz}.
    {\bf c} Transmon spectroscopy versus $\Phi_1$, at $B_r$~=~\SI{0}{mT} with ASQ1 open to its setpoint (see Tab.~\ref{tab:setpoints_new}) and ASQ2 closed. The two transmon frequencies correspond to the two possible states of ASQ1. 
    {\bf d} Transmon spectroscopy versus $\Phi_2$, at $B_r$~=~\SI{0}{mT} with ASQ2 open to its setpoint (see Tab.~\ref{tab:setpoints_new}) and ASQ1 closed. The two transmon frequencies correspond to the two possible states of ASQ2. For all panels, \Vc~=~\SI{1500}{mV}.
    The lines in {\bf c} and {\bf d} show the transmon transition spectrum given the spin-dependent part of the ASQ potentials found in {\bf a}, {\bf b} and the measured value of $E_{\rm c}$ (see Tab.~\ref{tab:circuit_parameters}). The lines in {\bf c} and {\bf d}  are best fits to the measured data at $\Phi_i$ being integer multiples of $\Phi_0/2$. From these transmon spectra, we estimate the values of the spin-independent Josephson energies $E^I_{\rm J, 1}/h$~=~\SI{1.79}{GHz} and $E^I_{\rm J, 2}/h$~=~\SI{0.53}{GHz}. 
    }
    \label{fig:parameter-extraction_new} 
\end{figure}

We start by performing basic characterization measurements. The values of the spin-independent, $E^I_{{\rm J}, i}$, and spin-dependent, $E^\sigma_{{\rm J}, i}$, Josephson energies for both qubits are extracted from spin-flip and transmon spectroscopy measurements (see Fig.~\ref{fig:parameter-extraction_new}). Fig.~\ref{fig:parameter-extraction_new}a and b show in-field spectroscopy of the ASQ1 and ASQ2 spin-flip transitions, respectively. The values of $E^\sigma_{\rm J, 1}$ and $E^\sigma_{\rm J, 2}$ are extracted from fits to a skewed (Eq.~\ref{eq:sin_sk}) and a non-skewed (Eq.~\ref{eq:sin}) sinusoidal relation, respectively. Fig.~\ref{fig:parameter-extraction_new}c and d show transmon spectroscopy measurements performed at zero magnetic field, each with only one of the two ASQs open (ASQ1 in panel c and ASQ2 in panel d). 
The values of $E^I_{{\rm J}, i}$ are estimated by fitting the measured transmon transitions with Eq.~\ref{Eq:total-transmon-Hamiltonian} at $\Phi_i$ being integer multiples of $\Phi_0/2$.
 
Before investigating the longitudinal coupling strength at the new gate setpoints, we characterize their magnetic field dependence. 
The characterization is done analogously to that for the previous gate setpoint (discussed around Fig.~\ref{fig:supplement-angle-dep}) and can be found in the data repository. The relevant extracted parameters are summarized in Tab.~\ref{tab:directions_new}. By performing spin-flip spectroscopy of each of the two qubits at different field directions, we extract their $g$-factors on the chip plane, which range between 6 and 15 for ASQ1 and between 9 and 15 for ASQ2. The values along the $B_z$ and $B_y$ directions are reported in Tab.~\ref{tab:directions_new}.
As before, we determine the spin-flip polarization direction for ASQ1 by finding two perpendicular directions in the $y-z$ and $x-z$ planes. The resulting spin-polarization direction is reported in Tab.~\ref{tab:directions_new} and is this time found to be approximately 1.85 degrees away from the nanowire axis. For ASQ2 we did not determine the full spin-orbit direction as we only data measured in the $y-z$ plane.

\begin{table*}[h!]
\begin{ruledtabular}
\begin{tabular}{cccccccc}
\textrm{}& $g_z^{{\rm ASQ}i}$ & $g_y^{{\rm ASQ}i}$  &  $\theta_{||}$  & $\phi_{||}$ & $\alpha$ \\
\colrule
ASQ1 & 14.9 & 6.7 & 1.85 & 64.4 & 1.36 \\
ASQ2 & 14.1 & - & - & - & - \\
\end{tabular}
\end{ruledtabular}
\caption{ {\bf Summary of $g$-factors and relevant angles for ASQ1 and ASQ2 at their alternative setpoint.}  
 The zero-field spin-polarization direction ($\theta_{||}, \phi_{||}$ in spherical coordinates) is calculated as the vector product of the two perpendicular directions $\theta_{yz, \perp}$  and  $\theta_{xz, \perp}$.
 $\alpha$ denotes the angle between the field applied for the longitudinal coupling measurements of Fig.~\ref{fig:fig_longitudinal_combined}
 and the spin-polarization direction for each qubit.
}
\label{tab:directions_new}
\end{table*}

\begin{figure}[h!]
\includegraphics{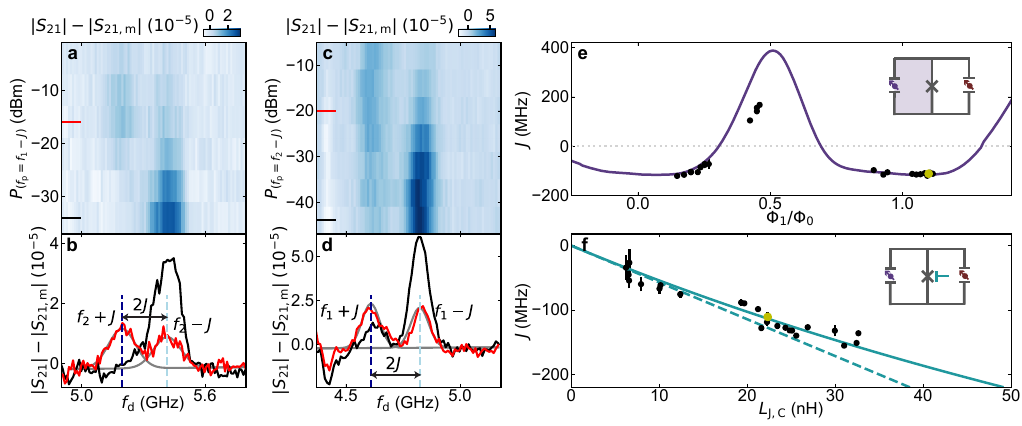}
\caption{ {\bf Longitudinal coupling between the two Andreev spin qubits at different gate setpoint.} 
Data measured and processed in the same way as for Figs.~3 and 4 in the main text, but now for alternative ASQ1 and ASQ2 gate setpoints, reported in Tab.~\ref{tab:setpoints_new}.
{\bf a} Spectroscopy of ASQ2 as a function of the drive frequency, $f_{\rm d}$ and the power of a pump tone resonant with ASQ1 at frequency $f_{\rm p} = f_1-J$.
{\bf b} Linecuts of {\bf a} at the powers indicated with color-matching lines. The grey line shows the result of a double-Gaussian fit to the signal in red (see Sec.~\ref{Ss:methods}) and the vertical lines indicate the best-fit values of the two Gaussian centers.
{\bf c}, {\bf d} Similar to the situation of {\bf a}, {\bf b}, but with the roles of ASQ1 and ASQ2 exchanged. In this case, the pump tone drives ASQ2 at a frequency $f_{\rm p}=f_2-J$, while performing spectroscopy of ASQ1. 
The longitudinal coupling strength, $J$, extracted from the fits is $-110.0\pm3.2$~MHz and $-107.0\pm3.1$~MHz, respectively for {\bf b} and {\bf d}.
For {\bf a} - {\bf d},  $\Phi_1=~0.1\Phi_0$.
{\bf e} Qubit-qubit coupling strength, $J$, as a function of flux in the loop for ASQ1, $\Phi_1$, see also inset. The values of $J$ are determined from a double Gaussian fit to a spectroscopy trace of one ASQ taken while driving the other ASQ.  The markers and error bars represent the best-fit values and their estimated standard errors (one-sigma confidence intervals), respectively. The purple line shows the expected dependence $hJ(\Phi_1)  = A \frac{L_{\rm J, C} L_{\rm ASQ}}{L_{\rm J, C}+L_{\rm ASQ}} I_{1}(\Phi_1) I_{2}/2$ for the value of $A$ extracted from panel {\bf f}.
{\bf f} Qubit-qubit coupling strength $J$ at fixed $\Phi_1=1.1 \Phi_0$ and as a function of $L_{\rm J, C}$, which is varied using the gate-voltage at the coupling junction (see inset). The continuous line shows a fit to the dependence $hJ(L_{\rm J, C})  = A \frac{L_{\rm J, C} L_{\rm ASQ}}{L_{\rm J, C}+L_{\rm ASQ}} I_{1} I_{2}/2$. We extract a value $A = 0.79 \pm 0.02$ from this fit.
The dashed line shows the linear dependence $hJ(L_{\rm J, C})= A L_{\rm J, C} I_{1} I_{2}/2$.
The yellow marker in {\bf e} and in {\bf f} is a shared point between the two panels. 
For all panels, $B_z=$~\SI{25}{mT}, $B_y=$~\SI{0.25}{mT},  $B_x=$~\SI{0.0}{mT} and $\Phi_2=0.48\Phi_0$.
}
\label{fig:fig_longitudinal_combined}
\end{figure}

Next, we measure the longitudinal coupling energy at the selected gate setpoints in the same way as for Fig.~3 in the main text. Fig.~\ref{fig:fig_longitudinal_combined}a-d show a longitudinal coupling measurement for fixed control parameters $B_z=$~\SI{25}{mT}, $\Phi_1=0.1\Phi_0$,  $\Phi_2=0.48\Phi_0$ and \Vc~=~\SI{180}{mV}. These parameters set $f_1=$~\SI{4.7}{GHz}, $f_2=$~\SI{5.3}{GHz} and $L_{\rm J, C} = $~\SI{22.3}{nH}. Similarly to Fig.~3 in the main text, we find that the frequency of each of the qubits splits by $2J$ when the other qubit is driven with a pump tone $f_{\rm p}$. From spectroscopy of ASQ2 while ASQ1 is driven (Fig.~\ref{fig:fig_longitudinal_combined}a, b), we find  $J=-110.0\pm3.2$~MHz, while from spectroscopy of ASQ1 while ASQ2 is driven (Fig.~\ref{fig:fig_longitudinal_combined}c, d), we find $-107.0\pm3.1$~MHz. These two values are equal within their one-sigma confidence intervals, consistent with the theory prediction.

Finally, we investigate the flux and $L_{\rm J, C}$ dependence of the coupling strength, similarly to how it is done in Fig.~4 of the main text. Fig.~\ref{fig:fig_longitudinal_combined}e, f show the tunability of $J$  at the setpoint of Tab.~\ref{tab:setpoints_new}. These measurements are taken at the same $B_z$ and $\Phi_2$ conditions as in Fig.~\ref{fig:fig_longitudinal_combined}a-d, which result on a fixed supercurrent difference through ASQ2, $I_{\rm 2}=$~\SI{-5.6}{nA}. Panel e shows the $\Phi_1$ dependence of $J$ at \Vc~=~\SI{180}{mV}, which fixes $L_{\rm J, C} = $~\SI{22.3}{nH}. Similarly to what was found in the main text, we observe that $J$ can take both positive and negative values 
and that it follows a similar shape as that predicted by Eq.~\ref{eq:J_current_derivatives}.
We however note that, especially around $\Phi_1=\Phi_0$,  the data deviates from the behavior predicted by Eq.~\ref{eq:J_current_derivatives}. This is due to the fact that this data is not taken for parameters consistent with the limit $L_{\rm J,C} \ll L_{{\rm J},i}^\sigma, L_{{\rm J},i}^I \forall i$. As shown in Fig.~\ref{fig:supplement-theory-realparameters}, when this limit is not met Eq.~\ref{eq:J_current_derivatives} overestimates the value of $J$ in the region of $\Phi_1 \sim \Phi_0$.

Finally, Fig.~\ref{fig:fig_longitudinal_combined}f shows the evolution of $J$ with  $L_{\rm J, C}$ at a fixed $\Phi_1=1.1 \Phi_0$ setpoint, indicated with a yellow marker in Fig.~\ref{fig:fig_longitudinal_combined}e, which sets $I_{\rm 1}=$~\SI{1.7}{nA}. As expected, we find an increase of the magnitude of $J$ with $L_{\rm J, C}$, which is proportional, with a scaling factor $A = 0.79 \pm 0.02$, to  Eq.~2 from the main text given $L_{\rm ASQ} = \frac{\Phi_0^2}{4\pi^2} / (E^I_{\rm J, 1}\cos(\frac{2\pi}{\Phi_0}\Phi_1)+E^I_{\rm J, 2}\cos(\frac{2\pi}{\Phi_0}\Phi_2))=$~\SI{176.9}{nH}.

\bibliography{bibliography_sup.bib}